\documentclass[prd,onecolumn,nopacs,nofootinbib]{revtex4}

\usepackage{amsmath}
\usepackage{graphicx}
\usepackage{amsfonts}
\usepackage{latexsym}
\usepackage{bbold}
\usepackage{wasysym}
\usepackage{calligra}
\usepackage{float}
\usepackage{ulem}
\usepackage{inputenc}
\usepackage{xspace}
\usepackage{url}
\usepackage{epstopdf}
\usepackage{tikz}

\newcommand{\be}{\begin{equation}}
\newcommand{\ee}{\end{equation}}
\newcommand{\beq} {\begin{equation}}
\newcommand{\eeq} {\end{equation}}
\newcommand{\ba}{\begin{eqnarray}}
\newcommand{\ea}{\end{eqnarray}}

\begin{document}

	\title{Exactly Solvable Connections in Metric-Affine Gravity}
	%\title{Scale-invariant and projective-invariant theories in metric-affine geometry}
	
	\author{Damianos Iosifidis}
	\affiliation{Institute of Theoretical Physics, Department of Physics
		Aristotle University of Thessaloniki, 54124 Thessaloniki, Greece}
	\email{diosifid@auth.gr}
	
	\date{\today}
	\begin{abstract}
		
		This article presents a systematic way to solve for the Affine Connection in Metric-Affine Geometry. We start by adding to the Einstein-Hilbert action, a general action that is linear in the connection and its partial derivatives and respects projective invariance. We then generalize the result for Metric-Affine $f(R)$ Theories. Finally, we generalize even further and add an action (to the Einstein-Hilbert) that has an arbitrary dependence on the connection and its partial derivatives. We wrap up our results as three consecutive Theorems. We then apply our Theorems  to  some simple examples in order to illustrate how the procedure works and also discuss the cases of dynamical/non-dynamical connections.
		
	\end{abstract}
	
	\maketitle
	
	\allowdisplaybreaks
	
	%\newpage
	
	\tableofcontents
	%\newpage
	
	\section{Introduction}
	\label{intro}

Geometrical modifications of Gravity by generalizing the affine connection have a long history and date back to the works of Weyl \cite{Weyl:1918ib} and Cartan \cite{cartan1922equations}. In Weyl's theory the connection was symmetric but not metric compatible while Cartan's was a metric one but with an antisymmetric part (torsion). A general space that has an affine connection that is neither metric compatible nor symmetric constitutes what is broadly known as non-Riemannian Geometry. The underlying  Gravity theory in such a geometry is called Metric-Affine Gravity\cite{hehl1995metric}. In the Metric-Affine formulation, the metric tensor $g_{\mu\nu}$ and the affine connection $\Gamma^{\lambda}_{\;\;\;\mu\nu}$ are treated as independent variables and a relation among them may be found only after using the field equations. In the general formulation, both the gravity and matter sectors can depend on the affine connection. The additional contributions in the Metric-Affine theories come from torsion and non-metricity. Torsion is the antisymmetric part of the connection and the non-metricity measures the failure of the connection to be metric compatible (see definitions in next chapter). Both of these features can be computed once an affine connection $\Gamma^{\lambda}_{\;\;\;\mu\nu}$ is given\footnote{To be more specific, this is true only for torsion. In order to compute the non-metricity tensor one also needs to have a metric (along with the affine connection).}.

Metric-Affine Theories of Gravitation are particularly interesting for studying modifications of Gravity (beyond General Relativity) because the modifications, in this case, are introduced naturally by extending the geometry to be non-Riemannian. In view of this, along with the need to modify General Relativity, the latter have attracted some attention during the past few years \cite{vitagliano2011dynamics,olmo2011palatini,sotiriou2007metric,vitagliano2010dynamics,olmo2009dynamical,sotiriou2010f}, especially when it comes to Palatini $f(R)$ Gravity \cite{olmo2011palatini,sotiriou2007metric}. The Palatini approach is based on the assumption that the matter part of the action does not depend on the connection. With such a simplifying assumption, it can be shown (see for instance  \cite{sotiriou2009f}) that the connection in Palatini $f(R)$ lacks dynamics and can be expressed in terms of the metric, its derivatives and the matter fields. The situation changes radically when one allows matter to couple to the connection. In this case (Metric-Affine $f(R)$) the connection becomes dynamical in general \cite{vitagliano2011dynamics}. Staying in the realm of Palatini Gravity it was shown in   \cite{allemandi2004accelerated}   (and also in \cite{olmo2009dynamical}) that for Ricci squared families of the type  $f(R,R_{(\mu\nu)}R^{(\mu\nu)})$ the affine connection can still be algebraically eliminated and carries no dynamics. The way to solve for the affine connection was also presented there \cite{allemandi2004accelerated,olmo2009dynamical}. This is not the case however when one generalizes to families of the type $f(R,R_{\mu\nu}R^{\mu\nu})$ and in this case the connection becomes dynamical, as shown in \cite{vitagliano2010dynamics}, even for the simplifying case of vanishing torsion. From an effective field theory perspective, theories containing second order invariants of torsion and non-metricity were studied\footnote{The renormalizability of theories containg quadratic torsion and non-metricity scalars was studied in \cite{pagani2015quantum}.} in \cite{vitagliano2014role} and \cite{vitagliano2011dynamics} where it was found that to this order the connection lacks dynamics, but of course will become dynamical once higher order terms are added.

 Therefore, from the above discussion we see that it is important to have a tool for obtaining the form of the affine connection for a given theory and see whether the latter becomes dynamical or not.
 It is the purpose of this article to present a systematic way to do so for specific  Metric-Affine theories. Also, since many families of the theories in this formalism share what is known as projective invariance\footnote{For a general discussion on the possible scale transformations in Metric-Affine Geometry see \cite{iosifidis2018scale}.} we also touch upon projective invariance breaking in Metric-Affine $f(R)$ theories. In particular, we review the two methods that have been suggested in the literature (\cite{1981GReGr..13.1037H,sotiriou2007metric})  in order to break this invariance and also present another possibility.

The paper is organized as follows. First, we introduce the basic ingredients that constitute the generalized geometry and discuss in some detail the geometrical meaning of torsion and non-metricity with some illustrative examples. The reader who is familiar with these concepts may skip this section. Then we present and prove step by step a systematic way to solve for the affine connection, firstly for theories of specific form and then later we generalize our result to $f(R)$ actions and finally to theories that have an arbitrary dependence on the connection. We state and prove our results as three consecutive Theorems. We then present an application of our first theorem in a simple torsion-full model. Then, we apply our derived results for the connection from Theorem-2, to Metric-Affine $f(R)$ Gravities and also touch upon projective invariance breaking in these theories. Finally we give an example of a theory with a dynamical connection (which falls in the category of our Theorem-3) and discuss the cases of dynamical/non-dynamical connections.
\newline\\
\textbf{Note:} In this report strong  emphasis is given on the mathematical procedures that take place in order to help the reader who is not familiar with these concepts, keep up with the discussion. The reader who is familiar enough with these ideas may skip the proceeding sections and  head directly to section VIII-(Exactly Solvable Connections) where the main results of this paper are presented.

\section{Introduction to Non-Riemannian geometry}
Let us introduce here the basic mathematical quantities that constitute a generalized non-Riemannian geometry. The most general Gravity Theory that is based on a non-Riemannian geometry is the so called Metric-Affine Gravity\cite{hehl1995metric}. First of all note that the term non-Riemannian refers to a generalized geometry where apart from the curvature the space is also endowed with torsion (i.e. vectors rotate upon parallel transport and as a result infinitesimal parallelograms do not exist) and non-metricity (dot products and lengths of vectors are not preserved while moving on the manifold). It is important to stress out that curvature, torsion and non-metricity are different geometrical entities on their own and we can have the one without necessarily the others. For example, we may have a space that is metric and flat but has a non-vanishing torsion. This is the case in what is known as the teleparallel formulation of Gravity\cite{aldrovandi2012teleparallel}. In this formulation curvature and non-metricity are zero and gravity is due to torsion (see \cite{aldrovandi2010introduction} for instance). There also exists the symmetric teleparallel formulation \cite{nester1999symmetric,jimenez2018teleparallel}  where one has zero curvature and torsion but a non-vanishing non-metricity. A space with  zero torsion and non-metricity but non-vanishing  curvature is our familiar Riemannian space of General Relativity. A space that has all three vanishing will be a Euclidean  (or Minkowski) space. 

Note that the three aforementioned geometrical quantities can all be calculated when the two fundamental objects of a manifold are given, a metric $g_{\mu\nu}$ and a connection $\Gamma^{\lambda}_{\;\;\;\mu\nu}$. The former defines distances and angles between vectors and the latter defines parallel transfer of vectors (or tensor fields in general) on the manifold. In a general non-Riemannian space, these two quantities (metric and connection) are independent and only become  interrelated when further assumptions are made. For instance, when one assumes  a torsion-free and  metric-compatible connection, the resulting connection is uniquely defined in terms of the metric tensor and its derivatives and is the familiar Levi-Civita connection (see subsequent discussion). We now proceed by giving the basic definitions of the geometrical objects that built a non-Riemannian geometry.

\subsection{Connection and Riemann tensor}
We shall start with  the general definitions of the connection,  the Riemann and the torsion tensor. We should point out that these definitions do not need the existence of a metric. Let us firstly introduce a general connection $\Gamma^{\alpha}_{\;\;\;\mu\nu}$ which is used in order to define parallel transport (through covariant differentiation) of tensorial fields. For a general tensorial field of rank (n,m) one has
\begin{gather}
\nabla_{\mu}T^{\alpha_{1}\alpha_{2}...\alpha_{n}}_{\;\;\;\;\beta_{1}\beta_{2}...\beta_{m}}=\partial_{\mu}T^{\alpha_{1}\alpha_{2}...\alpha_{n}}_{\;\;\;\;\beta_{1}\beta_{2}...\beta_{m}}+\Gamma^{\alpha_{1}}_{\;\;\;\rho\mu}T^{\rho\alpha_{2}...\alpha_{n}}_{\;\;\;\;\beta_{1}\beta_{2}...\beta_{m}}+...+\Gamma^{\alpha_{2}}_{\;\;\;\rho\mu}T^{\alpha_{a}\alpha_{2}...\alpha_{n-1}\rho}_{\;\;\;\;\beta_{1}\beta_{2}...\beta_{m}}  \\ 
-\Gamma^{\rho}_{\;\;\;\beta_{1}\mu}T^{\alpha_{1}\alpha_{2}...\alpha_{n}}_{\;\;\;\;\rho\beta_{2}...\beta_{m}}-...-\Gamma^{\rho}_{\;\;\;\beta_{m}\mu}T^{\alpha_{1}\alpha_{2}...\alpha_{n}}_{\;\;\;\;\rho\beta_{2}...\beta_{m-1}\rho} \nonumber
\end{gather}
Notice that according to our definition the index $\mu$ that appears in the covariant derivative is placed at the very right of the connection.\footnote{Some authors define it the other way around. It is important to strictly stick to whichever definition one adopts, since this will have an impact on the definition of the Riemann tensor.} 
In particular, for a mixed rank-$(1,1)$ tensorial field
the following holds true
\begin{equation}
\nabla_{\mu}T^{\alpha}_{\;\;\beta}=\partial_{\mu}T^{\alpha}_{\;\;\beta}+\Gamma^{\alpha}_{\;\;\;\rho\mu}T^{\rho}_{\;\;\beta}-\Gamma^{\rho}_{\;\;\;\beta\mu}T^{\alpha}_{\;\;\rho}
\end{equation}
Contracting in $\alpha$, $\beta$ the  above ($i.e.$ forming the scalar quantity $T\equiv T^{\alpha}_{\;\;\alpha}$) we immediately conclude that
\begin{equation}
\nabla_{\mu}T=\partial_{\mu}T
\end{equation}
confirming that on scalars covariant differentiation reduces to partial one. Now regarding scalar densities\footnote{Recall that a scalar density-$\mathcal{P}$ of weight $w$ transforms as $\mathcal{P} \rightarrow \mathcal{P}^{'}=J^{w}\mathcal{P}$ under a general coordinate transformation $x \rightarrow x^{'}=f(x)$. Notice that  the Jacobian of the transformation reads $J\equiv \Big|\frac{\partial x}{\partial x'}\Big|$ according to our definition. As a result the determinant of the metric tensor and the square root of it, are scalar densities of weights $+2$ and $+1$ respectively! If one defines the Jacobian $J\equiv \Big|\frac{\partial x'}{\partial x} \Big|$ then the above weights are $-2$ and $-1$ respectively.}, it holds that
\beq
\nabla_{\mu}\mathcal{P}=\partial_{\mu}\mathcal{P}-w\Gamma^{\lambda}_{\;\;\;\lambda\mu}\mathcal{P}
\eeq
for a scalar density $\mathcal{P}$ of weight $w$. Also, for a  tensor density $\mathcal{T}^{\alpha_{1}...\alpha_{n}}_{\;\;\;\;\beta_{1}...\beta_{m}}$ of weight $w$ one has
\begin{gather}
\nabla_{\mu}\mathcal{T}^{\alpha_{1}...\alpha_{n}}_{\;\;\;\;\beta_{1}...\beta_{m}}=\partial_{\mu}\mathcal{T}^{\alpha_{1}...\alpha_{n}}_{\;\;\;\;\beta_{1}...\beta_{m}}+\Gamma^{\alpha_{1}}_{\;\;\;\lambda\mu}\mathcal{T}^{\lambda...\alpha_{n}}_{\;\;\;\;\beta_{1}...\beta_{m}}+...+\Gamma^{\alpha_{n}}_{\;\;\;\lambda\mu}\mathcal{T}^{\alpha_{1}...\lambda}_{\;\;\;\;\beta_{1}...\beta_{m}} \nonumber \\
-\Gamma^{\lambda}_{\;\;\;\beta_{1}\mu}\mathcal{T}^{\alpha_{1}...\alpha_{n}}_{\;\;\;\;\lambda...\beta_{m}}-...-\Gamma^{\lambda}_{\;\;\;\beta_{m}\mu}\mathcal{T}^{\alpha_{1}...\alpha_{n}}_{\;\;\;\;\beta_{1}...\lambda}-w\Gamma^{\lambda}_{\;\;\;\lambda\mu}\mathcal{T}^{\alpha_{1}...\alpha_{n}}_{\;\;\;\;\beta_{1}...\beta_{m}}
\end{gather}
Notice the appearance of the term $-w\Gamma^{\lambda}_{\;\;\;\lambda\mu}\mathcal{T}^{\alpha_{1}...\alpha_{n}}_{\;\;\;\;\beta_{1}...\beta_{m}}$ with regards to the definition of the covariant derivative of a tensor field $(n,m)$.

Let us proceed now by giving the Riemann tensor. Forming the commutator of two covariant derivatives and acting it on a vector $u^{\mu}$ we arrive at
\begin{equation}
[\nabla_{\alpha} ,\nabla_{\beta}]u^{\mu}=2\nabla_{[\alpha} \nabla_{\beta]}u^{\mu}=R^{\mu}_{\;\;\;\nu\alpha\beta} u^{\nu}+2 S_{\alpha\beta}^{\;\;\;\;\;\nu}\nabla_{\nu}u^{\mu}
\end{equation}
where
\begin{equation}
R^{\mu}_{\;\;\;\nu\alpha\beta}:= 2\partial_{[\alpha}\Gamma^{\mu}_{\;\;\;|\nu|\beta]}+2\Gamma^{\mu}_{\;\;\;\rho[\alpha}\Gamma^{\rho}_{\;\;\;|\nu|\beta]}
\end{equation}
is the so-called Riemann tensor and the horizontal bars around an index denote that this index is left out of the (anti)-symmetrization. In addition, it appears the torsion tensor $ S_{\alpha\beta}^{\;\;\;\;\;\nu}$ which is given by the antisymmetric part of the connection\footnote{Note that even though the connection is not a tensor the difference between two connections does behave as a tensor.}
\begin{equation}
S_{\alpha\beta}^{\;\;\;\;\;\nu} := \Gamma^{\nu}_{\;\;[\alpha\beta]}=\frac{1}{2}(\Gamma^{\nu}_{\;\;\alpha\beta}-\Gamma^{\nu}_{\;\;\beta\alpha})
\end{equation}
Alternatively, one may also define the torsion tensor by acting the anti-symmetrized double covariant derivative to  a scalar, namely
\beq
\nabla_{[\mu} \nabla_{\nu]}\phi = S_{\mu\nu}^{\;\;\;\;\lambda}\nabla_{\lambda}\phi
\eeq
for any scalar $\phi$.
We should point out that by the above definition of the Riemann tensor alone, the only symmetry that the latter possesses is antisymmetry in its last two indices. Further symmetries appear only after imposing a torsionless ($ S_{\alpha\beta}^{\;\;\;\;\;\nu}=0$) and a metric compatible  ($\nabla_{\alpha}g_{\mu\nu}=0$) connection. This allows one to form the following contractions
\begin{equation}
R^{\mu}_{\;\;\;\mu\alpha\beta}\;,\;\;R^{\mu}_{\;\;\;\nu\mu\beta} \;,\;\;R^{\mu}_{\;\;\;\nu\alpha\mu}
\end{equation}
Note that the last contraction above (third term) is up to a minus sign equal to the second one and need not be considered separately. This defines the Ricci tensor 
\begin{equation}
R_{\nu\beta} :=  R^{\mu}_{\;\;\;\nu\mu\beta} = 2\partial_{[\mu}\Gamma^{\mu}_{\;\;\;|\nu|\beta]}+2\Gamma^{\mu}_{\;\;\;\rho[\mu}\Gamma^{\rho}_{\;\;\;|\nu|\beta]}
\end{equation}
which, is not symmetric in $\nu,\beta$ in general. In addition, the very first contraction above defines a new tensor which is non-vanishing only when non-metricity is present ($\nabla_{\mu}g_{\alpha\beta}\neq 0$), and goes by the name homothetic curvature
\begin{equation}
\hat{R}_{\alpha\beta} := R^{\mu}_{\;\;\;\mu\alpha\beta}=2\partial_{[\alpha}\Gamma^{\mu}_{\;\;\;|\mu|\beta]}=\partial_{\alpha}\Gamma^{\mu}_{\;\;\;\mu\beta}-\partial_{\beta}\Gamma^{\mu}_{\;\;\;\mu\alpha}
\end{equation}
Note now that for the above considerations no metric is required. When the space is also endowed with a metric tensor there is a third independent contraction that can be formed
\begin{equation}
\check{R}^{\mu}_{\;\;\beta} =g^{\nu\alpha}R^{\mu}_{\;\;\;\nu\alpha\beta} \equiv 2 g^{\nu\alpha}\partial_{[\alpha}\Gamma^{\mu}_{\;\;\;|\nu|\beta]}+2 g^{\nu\alpha} \Gamma^{\mu}_{\;\;\;\rho[\alpha}\Gamma^{\rho}_{\;\;\;|\nu|\beta]}
\end{equation}
However, the Ricci scalar is still uniquely defined since\footnote{Of course the other scalar that we can form by contracting the homothetic curvature with the metric is automatically zero since the former is antisymmetric and the latter symmetric in their indices.}
\begin{equation}
\check{R}=\check{R}^{\alpha}_{\;\;\alpha}=R^{\alpha}_{\;\;\;\beta\mu\alpha}g^{\beta\mu}=-R^{\alpha}_{\;\;\;\beta\alpha\mu}g^{\beta\mu}=-R_{\beta\mu}g^{\beta\mu}=-R
\end{equation}

\section{Torsion tensor and related vectors}
As we have already seen, the torsion tensor is defined as
\begin{equation}
S_{\mu\nu}^{\;\;\;\;\;\lambda} := \Gamma^{\lambda}_{\;\;[\mu\nu]}
\end{equation}
with this at hand we can define two new quantities. The first one is obtained by contracting in $(\mu=\lambda)$,
\begin{equation}
S_{\mu} := S_{\mu\lambda}^{\;\;\;\;\;\lambda}
\end{equation}
which we shall call the torsion vector. The second is a pseudo-vector that comes about when contracting with the Levi-Civita symbol, namely (in $4-dim$ for instance)
\begin{equation}
\tilde{S}^{\mu} := \epsilon^{\mu\nu\rho\sigma}S_{\nu\rho\sigma}
\end{equation}

\subsection{Geometrical Meaning of torsion}

The effect of torsion on geometrical grounds reflects the inability to form infinitesimal parallelograms when the latter is present. In others words we cannot form small parallelograms by parallel transportation of one vector to the direction of the other and vice versa. The end result is a pentagon.  To see this consider two curves $\mathcal{C}:x^{\mu}=x^{\mu}(\lambda)$ and $\mathcal{\tilde{C}}:\tilde{x}^{\mu}=\tilde{x}^{\mu}(\lambda)$ with tangent vectors
\begin{equation}
u^{\mu}=\frac{dx^{\mu}}{d\lambda}
\;\;\;\; and\;\;\;\;
\tilde{u}^{\mu}=\frac{d\tilde{x}^{\mu}}{d\lambda}
\end{equation}
respectively. Now, let us $d\tilde{x}^{\mu}$-displace $u^{\alpha}$ along $\mathcal{\tilde{C}}$ to obtain $u^{'\alpha}$ which in first order is given by
\begin{equation}
u^{'\alpha}=u^{\alpha}+(\partial_{\mu}u^{\alpha})d\tilde{x}^{\mu} \label{toru}
\end{equation}
but since $u^{\alpha}$ is parallely transported along $\mathcal{\tilde{C}}$, it holds that
\begin{equation}
\frac{d\tilde{x}^{\mu}}{d\lambda}\nabla_{\mu}u^{\alpha}=0= \frac{d\tilde{x}^{\mu}}{d\lambda}\partial_{\mu}u^{\alpha}+\Gamma^{\alpha}_{\;\;\;\nu\mu}\frac{d\tilde{x}^{\mu}}{d\lambda}u^{\nu} \nonumber
\end{equation}
or
\begin{equation}
(\partial_{\mu}u^{\alpha})d\tilde{x}^{\mu}=-\Gamma^{\alpha}_{\;\;\;\nu\mu}u^{\nu}\tilde{u}^{\mu}d\lambda
\end{equation}
which when substituted back in $(\ref{toru})$ results in
\beq
u^{'\alpha}=u^{\alpha}-\Gamma^{\alpha}_{\;\;\;\nu\mu}u^{\nu}\tilde{u}^{\mu}d\lambda
\eeq
Doing the same job but now for a $dx^{\mu}$-displacement of $\tilde{u}^{a}$ along $\mathcal{C}$, we get
\beq
\tilde{u}^{'\alpha}=\tilde{u}^{\alpha}-\Gamma^{\alpha}_{\;\;\;\nu\mu}\tilde{u}^{\nu}u^{\mu}d\lambda= \tilde{u}^{\alpha}-\Gamma^{\alpha}_{\;\;\;\mu\nu}\tilde{u}^{\mu}u^{\nu}d\lambda
\eeq
Subtracting the latter two, it follows that
\beq
(\tilde{u}^{\alpha}+u^{'\alpha})-(u^{\alpha}+\tilde{u}^{'\alpha})=2 S_{\mu\nu}^{\;\;\;\;\;\alpha}\tilde{u}^{\mu}u^{\nu}d\lambda
\eeq
Notice now that for the infinitesimal parallelogram to exist, the vectors $(\tilde{u}^{\alpha}+u^{'\alpha})$ and $(u^{\alpha}+\tilde{u}^{'\alpha})$ should be equal and as it is clear from the above, this is not true in the presence of torsion. Defining the vector that shows this deviation as $V^{\alpha}d\lambda=(\tilde{u}^{\alpha}+u^{'\alpha})-(u^{\alpha}+\tilde{u}^{'\alpha})$ the latter can also be written as\footnote{This only holds true for small displacements in the directions of $\tilde{u}^{\mu}$ and $u^{\nu}$ which themselves are computed at the starting point of the path.}
\beq
V^{\alpha}=2 S_{\mu\nu}^{\;\;\;\;\;\alpha}\tilde{u}^{\mu}u^{\nu}
\eeq
which is the vector that measures how much the parallelogram has been deformed.

\subsection{Illustrative Example}

Let us examine now the role of torsion, with a simple two dimensional example. Consider a $2-dim$ Euclidean (i.e flat) space with vanishing non-metricity but with a non-vanishing torsion. Take the  familiar orthonormal vector  basis $\{\bold{e}_{i}\}$ , $\;i=1,2$ on the $xy$-plane. Next, consider the lines $C$ :$y=0$ and $\tilde{C}$ :$x=0$ with tangent vectors $\bold{u}=\bold{e}_{1}$ and $\bold{\tilde{u}}=\bold{e}_{2}$ respectively. Now, take the vector $\bold{\tilde{u}}=\bold{e}_{2}$ and parallel transport it along the line $C$  a parameter distance $\lambda_{1}=1$ to obtain $\bold{\tilde{u}'}$. Also,  parallel transport  $\bold{u}=\bold{e}_{1}$ along $\tilde{C}$ a parameter distance $\lambda_{2}=1$ to obtain $\bold{u'}$. The linking vector between the two is
\begin{equation}
V^{\alpha}=2 S_{\mu\nu}^{\;\;\;\;\;\alpha}\tilde{u}^{\mu}u^{\nu} \label{vv}
\end{equation}
as we have already seen,and depends solely on torsion. To see now how is torsion related to rotations, denote as $\theta$ the angle between $\bold{\tilde{u}'}$ and the $x$-axis and as $\phi$ the angle between the vector $\bold{u'}$ and the $y$-axis\footnote{Bear in mind that the resulting vectors $\bold{\tilde{u}'}$,$\bold{u'}$ retain the length of the initial vectors $\bold{u}$,$\bold{\tilde{u}}$  which lengths in our case are both equal to one. If non-metricity was present their lengths would also change under parallel transport. In this example, however, we consider only torsion in order to explore its geometrical meaning. }. Then, by means of elementary vector analysis we find
\begin{equation}
\bold{\tilde{u}'}=\cos{\theta}\bold{e}_{1}+\sin{\theta}\bold{e}_{2}
\end{equation}
and
\begin{equation}
\bold{u'}=\sin{\phi}\bold{e}_{1}+\cos{\phi}\bold{e}_{2}
\end{equation}
Also, it holds that
\begin{equation}
\bold{\tilde{u}}+\bold{u'}+\bold{V}=\bold{u}+\bold{\tilde{u}'}
\end{equation}
so that
\begin{equation}
\bold{V}= (1+\cos{\theta}-\sin{\phi})  \bold{e}_{1}   +(\sin{\theta}-1-\cos{\phi})\bold{e}_{2} \label{tt}
\end{equation}

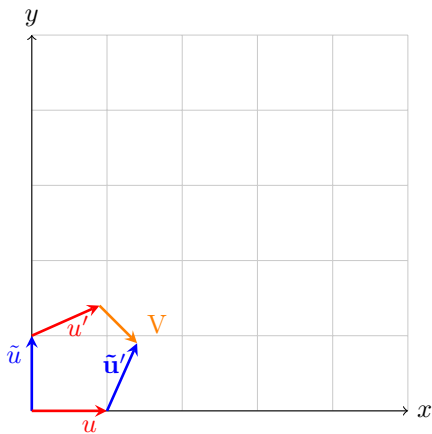
\begin{tikzpicture}
\draw[thin,gray!40] (0,0) grid (5,5);
\draw[->] (0,0)--(5,0) node[right]{$x$};
\draw[->] (0,0)--(0,5) node[above]{$y$};
\draw[line width=1pt,blue,-stealth](0,0)--(0,1) node[anchor=north east]{$\tilde{u}$};
\draw[line width=1pt,red,-stealth](0,0)--(1,0) node[anchor=north east]{$u$};
\draw[line width=1pt,blue,-stealth](1,0)--(1.4,0.9) node[anchor=north east]{$\bold{\tilde{u}'}$};
\draw[line width=1pt,red,-stealth](0,1)--(0.9,1.4) node[anchor=north east]{$u'$};
\draw[line width=1pt,orange,-stealth](0.9,1.4)--(1.4,0.9) node[anchor=south west]{V};

\end{tikzpicture}

Furthermore, using the fact that $u^{\mu}=\delta^{\mu}_{1}$ and $\tilde{u}^{\mu}=\delta^{\mu}_{2}$ equation ($\ref{vv}$) becomes
\begin{equation}
V^{\alpha}=2 S_{21}^{\;\;\;\;\alpha}
\end{equation}
or in components
\begin{equation}
V^{1}=2 S_{21}^{\;\;\;\;1}\;,\; \;V^{2}=2 S_{21}^{\;\;\;\;2}
\end{equation}
and by writing out $\bold{V}$ in the $\{\bold{e}_{i}\} $ basis
\begin{equation}
\bold{V}=V^{1}\bold{e}_{1} +V^{2}\bold{e}_{2}=2 S_{21}^{\;\;\;\;1}\bold{e}_{1} +2 S_{21}^{\;\;\;\;2}\bold{e}_{2}=S_{x}\bold{e}_{1}+S_{y}\bold{e}_{2}
\end{equation}
where we have defined $S_{x}\equiv 2 S_{21}^{\;\;\;\;1} $\;, $S_{y}\equiv 2 S_{21}^{\;\;\;\;2} $ the 2 only components of torsion in $2-dim$\footnote{Recall that in general $n-dim$ spaces the torsion tensor has $n^{2}(n-1)/2$ components.}. Comparing the above equation with ($\ref{tt}$) we find the relation between the components of torsion and the angles of rotation of the transported vectors
\begin{equation}
S_{x}=1+\cos{\theta}-\sin{\phi}
\end{equation}
\begin{equation}
S_{y}=\sin{\theta}-1-\cos{\phi}
\end{equation}
From these it is now pretty apparent how is torsion related to the rotation of vectors. Let us go one step further and compute the actual area of the pentagon that is formed due to torsion. Notice that if no torsion was present we would have the formation of a square (since we have picked $\lambda_{1}=\lambda_{2}=1$) with area $\sigma_{0}=1$ but now we have a pentagon and we would like to compute its area. One way to do this is by a specific application of Green's theorem which gives the area enclosed by a closed curve in terms of a closed line integral. As it is well known, it holds that 
\begin{equation}
\sigma =\oint_{C_{0}} x dy
\end{equation}
Breaking up the integral into its five individual line segments that constitute the pentagon we finally arrive at
\begin{gather}
\sigma ( \theta,\phi) =\frac{1}{2}\Big[ 2\cos{\theta} +\sin{\theta}\cos{\theta}-\sin{\phi}\cos{\phi}+\nonumber \\
(1+\sin{\phi}-\cos{\theta})(1+\sin{\theta}+\cos{\phi})\Big]
\end{gather}
After some rearranging, it can also be brought to the more symmetric form
\begin{gather}
\sigma ( \theta,\phi) =\frac{1}{2}\Big[ 1+\cos{\theta}+\cos{\phi}+\sin{\theta}+\sin{\phi}-\cos{(\theta +\phi)} \Big]
\end{gather}
and this is the area of the pentagon that did not close to square due to torsion. Notice that when there is no rotation (i.e torsion is zero) $\theta=0=\phi$ and $\sigma(0,0)=1$ the area of the square. Now, in the case where the effect of torsion is small, one can approximate $\sin{x}\simeq x$ and $\cos{x}\simeq 1$ where $x\ll 1$ stands for both $\theta,\phi$ such that $S_{x}\simeq\theta$\;, \; $S_{y}\simeq-\phi$ and the pentagon area is given by
\begin{gather}
\sigma ( \theta,\phi) \simeq 1+\frac{\theta+\phi}{2}
\end{gather}
or
\begin{gather}
\sigma ( \theta,\phi) \simeq 1+\frac{S_{x}-S_{y}}{2}=1+S_{21}^{\;\;\;\;1}+S_{12}^{\;\;\;\;2}
\end{gather}
in terms of the torsion components. Again, the unity on the right hand side is the area of the square that is formed when there is no torsion, and the rest is the modification of the original area due to torsion effects.

\section{Non-metricity Tensor and related vectors}
In a general metric affine space, as we have already pointed out, the connection is not metric compatible. This failure of the connection to covariantly conserve the metric is called non-metricity  and is defined as
\begin{equation}
Q_{\alpha\mu\nu} := -\nabla_{\alpha}g_{\mu\nu} \label{p}
\end{equation}
We should also mention that non-metricity is a quantity that depends both on the metric tensor and the connection. Indeed, expanding (\ref{p}) we obtain
\begin{equation}
Q_{\alpha\mu\nu} := -\nabla_{\alpha}g_{\mu\nu} =-\partial_{\alpha}g_{\mu\nu}+\Gamma^{\rho}_{\;\;\;\mu\alpha}g_{\rho\nu}+\Gamma^{\rho}_{\;\;\;\nu\alpha}g_{\mu\rho}
\end{equation}
from which, the dependence on $\Gamma^{\lambda}_{\;\;\;\mu\nu}$ and $g_{\mu\nu}$ is apparent. The corresponding expression for the non-metricity with upper indices is given by
\begin{gather}
Q_{\rho}^{\;\;\;\alpha\beta} := g^{\mu\alpha}g^{\nu\beta}Q_{\rho\mu\nu}=+\nabla_{\rho}g^{\alpha\beta}   \label{ol}
\end{gather}
which is easily verified by employing Leibniz's rule. Notice also the sign difference compared to the expression $(\ref{p})$. Having defined the non-metricity tensor there exist two independent vectors that one can form out of it. The first one is formed by contracting the second and third indices of the latter with the metric tensor  and goes by the name Weyl vector\footnote{In the literature it is common to also divide this vector by the spacetime dimensionality. That is $Q_{\mu}\rightarrow  Q_{\mu}/n$. However, our definition here does not include this factor.}
\begin{equation}
Q_{\alpha}:= g^{\mu\nu}Q_{\alpha\mu\nu}=Q_{\alpha\mu}^{\;\;\;\;\;\mu}=Q_{\alpha\;\;\;\;\mu}^{\;\;\;\mu}
\end{equation}
The second vector is formed by contracting the first and second indices with the metric\footnote{Note that the possibility to contract first and third index also exists. However, since non-metricity is symmetric in the second and third indices this vector would be the same with the one formed here.}, namely
\begin{equation}
\tilde{Q}_{\nu} := g^{\mu\alpha}Q_{\alpha\mu\nu}=Q^{\mu}_{\;\;\;\mu\nu}=-g^{\mu\alpha}\nabla_{\alpha}g_{\mu\nu}
\end{equation}
and does not seem to go with any particular name in the literature. We shall call it $2^{nd}$ non-metricity vector. We should point out that this is the same vector that one can form by contracting ($\ref{ol}$) in $\rho$ and $\alpha$ (or $\rho$ and $\beta$). Indeed, one has
\begin{equation}
\tilde{Q}^{\beta}:= Q_{\alpha}^{\;\;\;\alpha\beta}=\nabla_{\alpha}g^{\alpha\beta}=g^{\nu\beta}g^{\mu\alpha}Q_{\alpha\mu\nu}=g^{\nu\beta}\tilde{Q}_{\nu}
\end{equation} 
Thus, two independent vectors can be formed out of non-metricity and metric tensor alone.

\subsection{Geometrical meaning of Non-Metricity}

To see the effect on non-metricity in the space let us consider two vectors $a^{\mu}$ and $b^{\mu}$ and form their inner product $a\cdot b=a^{\mu}b^{\nu}g_{\mu\nu}$. Now, let us parallel transport both vectors along a given curve $\mathcal{C}: x^{\mu}=x^{\mu}(\lambda)$. For a Riemannian space (both torsion and non-metricity vanish) we know that upon such a transportation their inner product does not change, that is 
\beq
\frac{D}{d\lambda}(a\cdot b)=0
\eeq
When non-metricity is present a computation now reveals
\beq
\frac{D}{d\lambda}(a\cdot b)=\frac{dx^{\alpha}}{d\lambda}(\nabla_{\alpha} a^{\mu})b_{\mu}+\frac{dx^{\alpha}}{d\lambda}(\nabla_{\alpha} b^{\nu})a_{\nu}+\frac{dx^{\alpha}}{d\lambda}(\nabla_{\alpha}g_{\mu\nu})a^{\mu}b^{\nu}
\eeq
Now, since $a^{\mu}$ and $b^{\mu}$ are parallel transported along the curve, it holds that
\beq
\frac{dx^{\alpha}}{d\lambda}(\nabla_{\alpha} a^{\mu})=0\;,\;\;\frac{dx^{\alpha}}{d\lambda}(\nabla_{\alpha} b^{\nu})=0
\eeq
so we are left with
\beq
\frac{D}{d\lambda}(a\cdot b)=-Q_{\alpha\mu\nu}\frac{dx^{\alpha}}{d\lambda}a^{\mu}b^{\nu}
\eeq
from which we conclude that, when non-metricity is present, the inner product of two vectors does change when we parallel transport them along a curve. Note that for $b^{\mu}=a^{\mu}$ the above becomes
\beq
\frac{D}{d\lambda}(\|a \|^{2} )=-Q_{\alpha\mu\nu}\frac{dx^{\alpha}}{d\lambda}a^{\mu}a^{\nu} \label{fixedlvq}
\eeq
which means that the magnitude of a vector changes when we parallel transport it along a given curve! Therefore non-metricity has to do with vectors non-preserving their magnitudes and inner products.

\subsection{An illustrative example}
Let us find how does the length of a vector change in the case where the non-metricity is Weyl non-metricity. Recall that for Weyl geometry, we have
\beq
Q_{\alpha\mu\nu}=\frac{1}{n}Q_{\alpha}g_{\mu\nu}
\eeq 
and the length of a vector $a^{\mu}$, when transfered along a given curve $C:$ $x^{\alpha}=x^{\alpha}(\lambda)$, satisfies
\beq
\frac{D}{d\lambda}(\|a \|^{2} )=-\frac{1}{n}Q_{\alpha}\frac{dx^{\alpha}}{d\lambda}g_{\mu\nu}a^{\mu}a^{\nu}=-\frac{1}{n}Q_{\alpha}\frac{dx^{\alpha}}{d\lambda}\|a \|^{2}
\eeq
Setting $l^{2}=\|a \|^{2}$ and integrating that last one, it follows that
\beq
l(x)=l_{0}e^{-\frac{1}{2n}\int_{c} Q_{\alpha}dx^{\alpha}}
\eeq
from which we see that the change of the length is generally path dependent. In the case where the Weyl vector is exact, that is $Q_{\mu}=\partial_{\mu}\phi$ , we have what is known as a Weyl integrable geometry\footnote{For non-metricity of the generic form, the non-integrability is given the symmetric (in the first two indices) Riemann tensor (see \cite{iosifidis2018raychaudhuri} for instance for the identity relating this symmetric part with non-metricity). A special case of the latter being the homothetic curvature, whose vanishing gives the Weyl integrability condition. I am thankful to Tomi S. Koivisto for bringing this to my attention.} (WIG) for which the change on the vector's length depends only on the endpoints of the curve $C$, and for a closed loop the vector retains its initial length.

\subsection{Geometric Meaning of Homothetic Curvature}
Recall, that in a previous section we defined the homothetic curvature tensor  $\hat{R}_{\mu\nu}$ as the first contraction of the Riemann tensor $\hat{R}_{\mu\nu} := R^{\alpha}_{\;\;\;\alpha\mu\nu}$. This tensor has  a purely non-metric nature and is in fact related to the Weyl vector through
\beq
\hat{R}_{\mu\nu}=\frac{1}{2}(\partial_{\mu}Q_{\nu}-\partial_{\nu}Q_{\mu})=\partial_{[\mu}Q_{\nu]}
\eeq 
as can be easily checked. That is, the homothetic curvature is the curl of the Weyl vector. To see its geometrical meaning, let us go back to the length change of a vector when transfered along a curve $C$. If  $C$ is taken to be a closed curve (loop) then the length varies as
\beq
l(x)=l_{0}e^{-\frac{1}{2n}\oint_{c} Q_{\alpha}dx^{\alpha}} \label{leng}
\eeq
where the non-metricity was taken to be of Weyl type. Now, applying Stoke's theorem we have
\beq
\oint_{c} Q_{\alpha}dx^{\alpha}=\iint_{S}\partial_{[\mu}Q_{\nu]}d S^{\mu\nu}=\iint_{S}\hat{R}_{\mu\nu} d S^{\mu\nu}
\eeq
where $S$ is a surface that is enclosed by $C$ and $d S^{\mu\nu}$ the differential area element. Using this ($\ref{leng}$) becomes
\beq
l(x)=l_{0}e^{-\frac{1}{2 n}\iint_{S}\hat{R}_{\mu\nu} d S^{\mu\nu}} \label{hom}
\eeq
and from this we see that homothetic curvature is related with the length change that a vector experiences when transported along a closed loop. If non-metricity is weak, or the loop is small enough, by Taylor expanding we see that the total length change is given by
\beq
\delta l \simeq  -\frac{l_{0}}{2 n}\iint_{S}\hat{R}_{\mu\nu} d S^{\mu\nu}
\eeq
from which we see that the homothetic curvature serves as a generator of length changes of vector fields along closed paths.

\subsection{Toy Model}
Having established ($\ref{hom}$) let us play a little bit with the form of non-metricity to arrive at an interesting formula. To be more specific, consider a flat Euclidean $3-dim$ space that may posses non-vanishing non-metricity as well as torsion\footnote{The presence of torsion does not modify anything here, it simply rotates the vector when it is parallely transported along the curve. So, torsion rotates the vectors and non-metricity changes their lengths!}. Furthermore, assume we have a non-metric configuration with a non-metricity vector such that
\beq
\bold{Q}=\frac{3}{\alpha}(y \bold{e}_{1}-x \bold{e}_{2})  
\eeq
where $\alpha$ is a constant with area dimensions and $\bold{e}_{i}$, $\;i=1,2$ the usual orthonormal basis on the $xy$-plane. Take now the closed curve to lie  on the $xy$-plane, then
\begin{gather}
\oint_{c} Q_{\alpha}dx^{\alpha}=\oint_{c} \bold{Q} \cdot d\bold{r}=\iint_{S} (\bold{\nabla}\times \bold{Q})\cdot d \bold{S} = \nonumber \\
=-\frac{6}{\alpha}\iint _{S} d\sigma =-\frac{6}{\alpha} \sigma
\end{gather}
where $\sigma$ is the area enclosed by $C$. Substituting this back to ($\ref{leng}$) and setting $n=3$, we get
\beq
l(x)=l_{0}e^{-\frac{1}{6}\oint_{c} Q_{\alpha}dx^{\alpha}}=l_{0}e^{\frac{\sigma}{\alpha}}  \nonumber
\eeq
or
\beq
l(x)=l_{0}e^{\frac{\sigma}{\alpha}} 
\eeq
Thus, for such an arrangement of non-metricity the change in length of a vector transported along a closed curve $C$ depends on the surface area that $C$ encloses! In addition, if the ratio $\sigma/\alpha$ is small enough, the total change in length is exactly proportional to that surface, namely
\beq
\delta l \simeq \frac{l_{0}}{\alpha}\sigma
\eeq

\subsection{Fixed Length Vectors}

Now as we have seen, one consequence of non-metricity is that it changes the length of the vectors\footnote{The other consequence is the change of the dot product of two vectors.} when we transport them in space. So, one may ask are their any vectors, that retain their length in the presence of non-metricity? For generic non-metricity the answer is no. However, there exists a type of non-metricity for which we have vectors that remain unchanged. These are called $\bold{fixed}$ $\bold{length}$ $\bold{vectors}$. To see what kind of non-metricity allows for the existence of such vectors let us have a careful look at $(\ref{fixedlvq})$,
\beq
\frac{D}{d\lambda}(\|a \|^{2} )=-Q_{\alpha\mu\nu}\frac{dx^{\alpha}}{d\lambda}a^{\mu}a^{\nu}
\eeq
Taking $a^{\mu}$ to be proportional to $dx^{\mu}/d\lambda$ we obtain
\beq
\frac{D}{d\lambda}(\|a \|^{2} )\propto -Q_{\alpha\mu\nu}a^{\alpha}a^{\mu}a^{\nu}=-Q_{(\alpha\mu\nu)}a^{\alpha}a^{\mu}a^{\nu}
\eeq
Form the above we see that in order to have fixed lengths the right hand side must be zero, and given that $a^{\mu}$ is random we must have
\beq
Q_{(\alpha\mu\nu)}=0 \label{flnmt}
\eeq
in order for the theory to possess fixed length vectors. Any non-metricity that has vanishing totally symmetric part will admit fixed length vectors. This condition is also presented in the classic Schroendinger's $Spacetime-Structure$ \cite{schrodinger1985space}.  Let us go one step further and actually compute the simplest form of such non-metricity. The most straightforward decomposition of such a tensor would be in terms of a vector field, say $v_{\mu}$ and the metric $g_{\mu\nu}$, so that
\beq
Q_{\alpha\mu\nu}= a v_{\alpha}g_{\mu\nu}+b g_{\alpha(\mu}v_{\nu)}+c v_{\mu}v_{\nu}v_{\alpha}
\eeq
where $a,b,c$ are parameters to be computed and we demanded that the combinations are symmetric in $\mu,\nu$. Now since $Q_{\alpha\mu\nu}$ cannot have a totally symmetric part the last term on the right hand side of the above must be absent, and hence $c=0$. Now, demanding $Q_{(\alpha\mu\nu)}a^{\alpha}a^{\mu}a^{\nu}=0$ for random $a^{\mu}$ we get the relation $a= -b$. Notice also that we may set $a=1$ since this $a$ can be absorbed in a redefinition of $v_{\mu}$. Taking all the above into consideration, we finally arrive at
\beq
Q_{\alpha\mu\nu}=  v_{\alpha}g_{\mu\nu}-g_{\alpha(\mu}v_{\nu)}
\eeq
 and we can easily check that this form of non-metricity indeed satisfies $Q_{(\alpha\mu\nu)}=0$. 
Now, as can be easily checked by contracting with the metric tensor, the Weyl and second non-metricity vectors, are related to this $v_{\mu}$ through
\beq
Q_{\mu}=(n-1)v_{\mu}\;,\;\; \tilde{Q}_{\mu}=-\frac{(n-1)}{2}v_{\mu}
\eeq
From which we establish the relation between the two non-metricity vectors
\beq
Q_{\mu}=-2 \tilde{Q}_{\mu}
\eeq
Interestingly, this kind of non-metricity (that preserves lengths) overcomes Einstein's objection to the Weyl theory of unification\footnote{In Weyl's theory the non-metric tensor was given by $Q_{\alpha\mu\nu}=\frac{1}{n}Q_{\alpha}g_{\mu\nu}$ which definitely does not satisfy $Q_{(\alpha\mu\nu)}=0$ and therefore does not preserve the lengths of vectors.}. To recap, if the non-metricity is of of the form $(\ref{flnmt})$ the theory possesses fixed length vectors.

\section{Connection decomposition}
Having defined torsion and non-metricity we are now in a position to decompose the general connection in terms of the latter plus the Levi-Civita connection. To do so, we start by writing out the definition of the non-metricity 
\begin{equation}
Q_{\alpha\mu\nu}=-\nabla_{\alpha}g_{\mu\nu} =-\partial_{\alpha}g_{\mu\nu}+\Gamma^{\rho}_{\;\;\;\mu\alpha}g_{\rho\nu}+\Gamma^{\rho}_{\;\;\;\nu\alpha}g_{\mu\rho}
\end{equation}
and upon successive  permutations $\alpha\rightarrow \mu$, $\mu\rightarrow \nu$, $\nu\rightarrow \alpha$ on the above we may subtract $Q_{\mu\nu\alpha}$ and $Q_{\nu\alpha\mu}$ from the latter, we obtain  \footnote{We do assume that the metric tensor is symmetric since any antisymmetric part of it lacks a geometrical interpretation.}
\newline
\newline
\begin{gather}
-Q_{\alpha\mu\nu}+Q_{\mu\nu\alpha}+Q_{\nu\alpha\mu}=-(\partial_{\mu}g_{\nu\alpha}+\partial_{\nu}g_{\alpha\mu}-\partial_{\alpha}g_{\mu\nu}) \nonumber \\
+2\Gamma^{\rho}_{\;\;\;(\mu\nu)}g_{\alpha\rho}+2\Gamma^{\rho}_{\;\;\;[\alpha\nu]}g_{\mu\rho}+2\Gamma^{\rho}_{\;\;\;[\alpha\mu]}g_{\nu\rho}
\end{gather}
In addition, substituting
\begin{equation}
\Gamma^{\alpha}_{\;\;\;[\beta\gamma]}=S_{\beta\gamma}^{\;\;\;\;\alpha}
\end{equation}
we finally arrive at
\begin{gather}
\Gamma^{\lambda}_{\;\;\;\mu\nu}=\frac{1}{2}g^{\alpha\lambda}(\partial_{\mu}g_{\nu\alpha}+\partial_{\nu}g_{\alpha\mu}-\partial_{\alpha}g_{\mu\nu}) \nonumber \\
+\frac{1}{2}g^{\alpha\lambda}(Q_{\mu\nu\alpha}+Q_{\nu\alpha\mu}-Q_{\alpha\mu\nu}) -g^{\alpha\lambda}(S_{\alpha\mu\nu}+S_{\alpha\nu\mu}-S_{\mu\nu\alpha})
\end{gather}
We recognize the first part on the right-hand side as the Levi-Civita connection for which we use the tilde notation to distinguish it from the general connection, namely
\begin{equation}
\tilde{\Gamma}^{\lambda}_{\;\;\;\mu\nu}:=\frac{1}{2}g^{\alpha\lambda}(\partial_{\mu}g_{\nu\alpha}+\partial_{\nu}g_{\alpha\mu}-\partial_{\alpha}g_{\mu\nu})
\end{equation}
Thus,
\begin{equation}
\Gamma^{\lambda}_{\;\;\;\mu\nu}=\tilde{\Gamma}^{\lambda}_{\;\;\;\mu\nu}+\frac{1}{2}g^{\alpha\lambda}(Q_{\mu\nu\alpha}+Q_{\nu\alpha\mu}-Q_{\alpha\mu\nu}) -g^{\alpha\lambda}(S_{\alpha\mu\nu}+S_{\alpha\nu\mu}-S_{\mu\nu\alpha}) \label{affconection}
\end{equation}
we have fully decomposed the connection into a Riemannian-part (Levi-Civita connection), a contribution coming from non-metricity and another one due to torsion. It is common to introduce, at this point, a tensor which measures the deviation of the general connection with respect to the Levi-Civita one. This is the so-called $distortion$ tensor\footnote{Again, even though connections are not tensors,  the difference between connections defines $'legal'$ tensors.}
\begin{gather}
N^{\lambda}_{\;\;\;\;\mu\nu}:=\Gamma^{\lambda}_{\;\;\;\mu\nu}-\tilde{\Gamma}^{\lambda}_{\;\;\;\mu\nu}= \nonumber \\
\frac{1}{2}g^{\alpha\lambda}(Q_{\mu\nu\alpha}+Q_{\nu\alpha\mu}-Q_{\alpha\mu\nu}) -g^{\alpha\lambda}(S_{\alpha\mu\nu}+S_{\alpha\nu\mu}-S_{\mu\nu\alpha})
\end{gather} 
or
\beq
N_{\alpha\mu\nu}=\frac{1}{2}(Q_{\mu\nu\alpha}+Q_{\nu\alpha\mu}-Q_{\alpha\mu\nu}) -(S_{\alpha\mu\nu}+S_{\alpha\nu\mu}-S_{\mu\nu\alpha}) \label{l}
\eeq
In addition, the combination 
\begin{equation}
K_{\mu\nu}^{\;\;\;\;\lambda} :=g^{\alpha\lambda}(S_{\alpha\mu\nu}+S_{\alpha\nu\mu}-S_{\mu\nu\alpha})
\end{equation}
appearing above is oftentimes referred to as the $contorsion$. Note that we can split the distortion tensor into some symmetric and antisymmetric parts. Indeed, taking the symmetric part of (\ref{l}) in $\alpha, \mu$ and using the symmetries of $Q_{\alpha\mu\nu}$ and $S_{\alpha\mu\nu}$ we arrive at\footnote{Another way to derive this is by starting from the definition of non-metricity, (covariant derivative of the metric tensor )decompose the connection into the Levi-Civita and its non-Riemannian parts and use the fact that the non-metricity of the Levi-Civita connection is zero. }
\beq
Q_{\nu\alpha\mu}=2 N_{(\alpha\mu)\nu}
\eeq
While, when one takes the antisymmetric part in $\mu,\nu$ arrives at
\beq
S_{\mu\nu\alpha}=N_{\alpha[\mu\nu]}
\eeq
In addition, its totally antisymmetric part is given by
\beq
N_{[\alpha\mu\nu]}=S_{[\mu\nu\alpha]}=S_{[\alpha\mu\nu]}
\eeq
as can be easily checked. Note also that when we are looking at the autoparallels only the symmetric part   $N^{\lambda}_{\;\;\;\;(\mu\nu)}$ contributes to the equation, which is equal to
\beq
N^{\lambda}_{\;\;\;\;(\mu\nu)}=\frac{1}{2}g^{\alpha\lambda}(2 Q_{(\mu\nu)\alpha}-Q_{\alpha\mu\nu})-g^{\alpha\lambda}2 S_{\alpha(\mu\nu)}
\eeq
and from this, it is apparent that a completely antisymmetric torsion $(S_{\alpha\mu\nu}=S_{[\alpha\mu\nu]})$ has no effect on autoparallels.

\section{Energy-momentum and Hyper-momentum Tensors}
Having defined and briefly explored the generalized geometry let us continue by introducing the physical content that gives rise to such a geometry. Following the literature we define the
Energy-Momentum Tensor as the variation of the matter sector (of the action) with respect to the metric, namely
\beq
T_{\alpha\beta}:= -\frac{2}{\sqrt{-g}}\frac{\delta S_{M}}{\delta g^{\alpha\beta}}=-\frac{2}{\sqrt{-g}}\frac{\partial(\sqrt{-g} \mathcal{L}_{M})}{\partial g^{\alpha\beta}}
\eeq
Now, since matter can also depend on the affine connection, its variation with respect to it defines the  Hyper-momentum tensor \cite{hehl1976hypermomentum}
\beq
\Delta_{\lambda}^{\;\;\;\mu\nu}:= -\frac{2}{\sqrt{-g}}\frac{\delta S_{M}}{\delta \Gamma^{\lambda}_{\;\;\;\mu\nu}}=-\frac{2}{\sqrt{-g}}\frac{\partial ( \sqrt{-g} \mathcal{L}_{M})}{\partial \Gamma^{\lambda}_{\;\;\;\mu\nu}}
\eeq
An important thing that is almost never mentioned in the literature is that the above two tensors are not completely independent. Indeed, since $g_{\alpha\beta}$ and $\Gamma^{\lambda}_{\;\;\;\mu\nu}$ are independent variables, it holds that
\beq
\frac{\partial^{2}(\sqrt{-g} \mathcal{L}_{M})}{\partial g^{\alpha\beta}\partial \Gamma^{\lambda}_{\;\;\;\mu\nu}}=\frac{\partial^{2} (\sqrt{-g}\mathcal{L}_{M})}{\partial \Gamma^{\lambda}_{\;\;\;\mu\nu} \partial g^{\alpha\beta} }
\eeq
and as a result
\beq
\frac{1}{\sqrt{-g}}\frac{\partial}{\partial g^{\alpha\beta}}\Big( \sqrt{-g} \Delta_{\lambda}^{\;\;\;\mu\nu} \Big)= \frac{\partial T_{\alpha\beta}}{\partial \label{emhpt} \Gamma^{\lambda}_{\;\;\;\mu\nu}}
\eeq
Therefore we see that the energy-momentum and hyper-momentum tensors are not independent. If the latter is applied for a perfect fluid for instance, where $T_{\mu\nu}$ is independent of the connection, the hyper-momentum tensor has to satisfy
\beq
\frac{\partial}{\partial g^{\alpha\beta}}\Big( \sqrt{-g} \Delta_{\lambda}^{\;\;\;\mu\nu} \Big)=0
\eeq
So for a perfect fluid\footnote{Assuming that its form remains the same as in GR.}
\beq
\sqrt{-g} \Delta_{\lambda}^{\;\;\;\mu\nu}=independent \;of\;\; g_{\mu\nu}
\eeq
In addition, in the so-called Palatini Theories the matter action $S_{M}$ is assumed to be independent of the connection and therefore $\Rightarrow \Delta_{\lambda}^{\;\;\;\mu\nu} =0$. The latter means that in this case (Palatini Gravity) the energy momentum tensor is independent of the connection, as seen from $(\ref{emhpt})$. This result is crucial when studying the dynamical content of a connection and we will use it latter on when we touch upon the subject of dynamical/non-dynamical connections.

\section{Einstein's Theory in the Metric-Affine Framework}
As a warm up, we will show the known result that starting with the Einstein-Hilbert action and no matter fields, one ends up with Einstein Gravity plus an additional unspecified vectorial degree of freedom that gives rise to both torsion and non-metricity. This degree of freedom, however, can be eliminated by means of a projective transformation of the connection. This is possible because of the projective invariance of the Ricci scalar. However, this invariance is the very reason that renders the field equations problematic when one tries to add to the model a matter action that depends both on the metric and the connection. Then, one arrives at inconsistent field equations\footnote{This inconsistency arises due to the invariance of the Ricci scalar under projective transformations of the connection as we have already pointed out and is expressed as an unphysical constraint imposed on the matter fields. Note however that these constraints may not be so 'unphysical' and in certain cases may be even desirable as argued in \cite{jimenez2017born}(see also subsequent discussion).}. This inconsistency can be handled by fixing to zero the vector components of either the torsion or Weyl vectors but it seems that the situation suggests that in the MAG framework more general actions than the Einstein-Hilbert (or general $f(R)$) should be used. However, staying in the case of $f(R)$ we will review the two proposed ways to break projective invariance and also propose another possibility but only after presenting the way to solve for the affine connection (our three promised Theorems).

\subsection{Vacuum Einstein's Theory in MAG}
Let us start with the Einstein-Hilbert action in $n$-dimensions\footnote{Note that when torsion and non-metricity are present, the Ricci tensor is not symmetric. However, when contracted with the metric tensor to obtain the Ricci scalar, only the symmetric part is involved since the metric, being symmetric, symmetrizes everything contracted to it. In other words, the antisymmetric part of $R_{\mu\nu}$ drops out when constructing $R$.}
\begin{equation}
S_{EH}[g_{\mu\nu},\Gamma^{\lambda}_{\;\;\;\alpha\beta}]=\int d^{n}x\sqrt{-g}R=\int d^{n}x\sqrt{-g}g^{\mu\nu}R_{\mu\nu} =\int d^{n}x\sqrt{-g}g^{\mu\nu}R_{(\mu\nu)} \label{eihh}
\end{equation}
and no matter fields. Here, no a priori relation between the metric tensor $g_{\mu\nu}$ and the connection $\Gamma^{\lambda}_{\;\;\;\alpha\beta}$ has been assumed and therefore we have not assumed any torsionlessness and metric compatibility of the connection to begin with. Varying ($\ref{eihh}$) with respect to $g_{\mu\nu}$ and  $\Gamma^{\lambda}_{\;\;\;\mu\nu}$ and recalling that $R_{\mu\nu}$ is independent of the metric, we derive 
\begin{equation}
 R_{(\mu\nu)}-\frac{g_{\mu\nu}}{2}R=0
\end{equation}
\begin{equation}
-\nabla_{\lambda}(\sqrt{-g}g^{\mu\nu})+\nabla_{\sigma}(\sqrt{-g}g^{\mu\sigma})\delta^{\nu}_{\lambda} \\
+2\sqrt{-g}(S_{\lambda}g^{\mu\nu}-S^{\mu}\delta_{\lambda}^{\nu}+g^{\mu\sigma}S_{\sigma\lambda}^{\;\;\;\;\nu})=0 \label{nmh}
\end{equation}
the last one is the equation that relates the metric tensor and the connection. It is common in the literature to denote the left hand side of the above equation (divided by $\sqrt{-g}$) as $P_{\lambda}^{\;\;\;\mu\nu}$ and call it the Palatini tensor. In words
\beq
 P_{\lambda}^{\;\;\;\mu\nu}=-\frac{\nabla_{\lambda}(\sqrt{-g}g^{\mu\nu})}{\sqrt{-g}}+\frac{\nabla_{\sigma}(\sqrt{-g}g^{\mu\sigma})\delta^{\nu}_{\lambda}}{\sqrt{-g}} \\
+2(S_{\lambda}g^{\mu\nu}-S^{\mu}\delta_{\lambda}^{\nu}+g^{\mu\sigma}S_{\sigma\lambda}^{\;\;\;\;\nu})
\eeq
Note that in the above case (Einstein-Hilbert action with no matter fields) the Palatini tensor has to vanish. The Palatini tensor has only $n(n^{2}-1)$ instead of $n^{3}$ due to the fact that is traceless 
\begin{equation}
P_{\mu}^{\;\;\;\mu\nu}=0
\end{equation}
which is a general property and kills off $n$-equations\footnote{That is, in $4$-dim the Palatini tensor has $60$ components while the remaining $4$ components cannot be specified because of its traceless property.}. This implies that a vectorial degree of freedom is left unspecified and as a result the connection can only be determined up to a vector or more precisely a one form.\footnote{This is so because the Ricci scalar is invariant under projective transformations of the connection $
	\Gamma^{\lambda}_{\;\;\;\mu\nu} \rightarrow \Gamma^{\lambda}_{\;\;\;\mu\nu} +\delta^{\lambda}_{\mu}\xi_{\nu}
	$
	where $\xi_{\nu}$ is an arbitrary vector field (or more appropriately a one form). The identities that follow for scalars that are invariant under projective, conformal and frame rescaling transformations can be found in \cite{iosifidis2018scale}.
}More specifically, as we prove in the appendix, equation ($\ref{nmh}$) implies that the connection takes the following form
\begin{equation}
\Gamma^{\lambda}_{\;\;\;\mu\nu}= \tilde{\Gamma}^{\lambda}_{\;\;\;\mu\nu} -\frac{2}{(n-1)}S_{\nu}\delta_{\mu}^{\lambda}=\tilde{\Gamma}^{\lambda}_{\;\;\;\mu\nu}+\frac{1}{2 n}\delta_{\mu}^{\lambda}Q_{\nu} \label{llg}
\end{equation}
where $\tilde{\Gamma}^{\lambda}_{\;\;\;\mu\nu}$ is the Levi-Civita connection. This result can also be easily obtained by using our first Theorem (see next section).
Therefore, we conclude that indeed the connection is determined only up to an unspecified vectorial degree of freedom. The above result has also been given and discussed in \cite{dadhich2012equivalence,bernal2017non}. This additional degree of freedom can be gauged away by means of a projective transformation of the connection
\begin{equation}
\Gamma^{\lambda}_{\;\;\;\mu\nu}\longrightarrow \Gamma^{\lambda}_{\;\;\;\mu\nu}+\delta_{\mu}^{\lambda}\xi_{\nu}
\end{equation}
if $\xi_{\nu}$ is chosen to be equal to -$Q_{\nu}/2n$. For an interpretation of this one-form $\xi_{\nu}$ and a further discussion on the subject of projective invariance we refer the reader to \cite{bernal2017non}.  In addition, for connections of the form of ($\ref{llg}$) only the Levi-Civita part contributes in both the Einstein-Hilbert action and Einstein's equations.  In the end, as we show in the appendix, we have
\begin{equation}
\tilde{R}_{\mu\nu}-\frac{1}{2}\tilde{R}g_{\mu\nu}=0
\end{equation}
\begin{equation}
Q_{\alpha\mu\nu}=\frac{1}{n}Q_{\alpha}g_{\mu\nu} \;,\;S_{\mu\nu}^{\;\;\;\;\lambda}=-\frac{2}{(n-1)}S_{[\nu}\delta_{\mu]}^{\lambda}\;,\;\;S_{\lambda}=-\frac{(n-1)}{4 n}Q_{\lambda}
\end{equation}
where tilded quantities represent Riemannian parts (i.e. computed with respect to the Levi-Civita connection).   Thus we see that both torsion and non-metricity are non vanishing and depend on an unspecified vectorial degree of freedom. This is a consequence of the projective invariance of the Einstein-Hilbert action (which results in the tracelessness of the Palatini tensor $P_{\mu}^{\;\;\;\mu\nu}=0$). We conclude therefore that the Einstein-Hilbert action (without any matter fields) in the Metric-Affine framework does not reproduce exactly Einstein's theory. What it gives is, Einstein field equations along with an additional vectorial degree of freedom that produces non-vanishing torsion and non-metricity. However, these degrees of freedom are absent from the Einstein field equations. Having reviewed this classic result, let us now present the three Theorems for the connection expression.

\section{Exactly Solvable Models/Solving for the Affine Connection}
Let us  now give a systematic way to solve for the affine connection in Metric-Affine Theories. We state and prove our results as three subsequent Theorems. First we start by allowing actions that are linear in the connection to be added to the Einstein Hilbert. The expression for the connection is then given by Theorem-1. Then, in Theorem-$2$ we generalize for $f(R)$ and in the last case we assume no restriction on the additional part of the action (Theorem-3). We then see some applications of our results with three simple examples and discuss the conditions for obtaining dynamical/non-dynamical connections.
\subsection{Expression for an Exactly Solvable Connection}
Let us start with our first Theorem\footnote{Here we will follow a step by step proof, in order to make the procedure of solving with respect to affine connection completely clear, since we think that such a  systematic procedure is absent from the literature.}
\newline \\
\textbf{Theorem 1:} Consider the action
\begin{equation}
S[ g_{\mu\nu},\Gamma^{\lambda}_{\;\;\;\alpha\beta},\phi]=\frac{1}{2\kappa}\int d^{n}x\sqrt{-g}R+S_{1}[ g_{\mu\nu},\Gamma^{\lambda}_{\;\;\;\alpha\beta},\phi] \label{senar}
\end{equation}
where $\phi$ denotes any other additional fields that may be present in the space and 
\begin{equation}
S_{1}[ g_{\mu\nu},\Gamma^{\lambda}_{\;\;\;\alpha\beta},\phi]= \int d^{n}x\sqrt{-g} \mathcal{L}_{1}(g,\Gamma,\phi)
\end{equation}
Now given any general action $S_{1}[g,\Gamma,\phi]$ that is\footnote{Notice that we made no assumption about the origin of the action. It may include both matter and gravitational parts so long as it satisfies the requirements that we impose! However, a gravitational sector that is linear in the connection is difficult to come up with, we just include it for generality. On the contrary, a matter Lagrangian density linear in the connection is well motivated. For instance for a Dirac field in the presence of torsion one has $\mathcal{L}_{D}=\frac{i}{2}(\bar{\psi}\gamma^{\mu}D_{\mu} \psi -D_{\mu}\bar{\psi}\gamma^{\mu}\psi)-m \bar{\psi}\psi$ where the covariant derivative $D_{\mu}$ is linear in the contorsion and therefore linear in the connection as well. In the second Theorem we will assume that $S_{1}$ contains only a matter sector.} 
\begin{itemize}
	\item At most linear in $\Gamma^{\lambda}_{\;\;\;\mu\nu}$ and its partial derivatives
	\item Projective invariant
\end{itemize}
we state that the affine connection can solely be expressed in terms of variations of $\mathcal{L}_{1}$\footnote{Of course the result also contains the metric tensor and its derivatives as they appear for instance in the Levi-Civita part, but since this is too obvious we will omit mentioning it.} and its form is the following
\begin{equation}
\Gamma^{\lambda}_{\;\;\;\mu\nu}=\tilde{\Gamma}^{\lambda}_{\;\;\;\mu\nu}-\frac{g^{\lambda\alpha}}{2}(B_{\alpha\mu\nu}-B_{\nu\alpha\mu}-B_{\mu\nu\alpha})-\frac{g^{\alpha\lambda}}{(n-2)}g_{\nu[\mu}(B_{\alpha]}-\tilde{B}_{\alpha]})
\end{equation}
where
\begin{equation}
B_{\lambda}^{\;\;\;\mu\nu}:=  \frac{2 \kappa}{\sqrt{-g}}\frac{\delta  S_{1}}{\delta \Gamma^{\lambda}_{\;\;\;\mu\nu}}=\frac{2 \kappa}{\sqrt{-g}}\frac{\partial (\sqrt{-g}\mathcal{L}_{1})}{\partial \Gamma^{\lambda}_{\;\;\;\mu\nu}}
\end{equation}
and $B^{\mu}:= B_{\lambda}^{\;\;\;\mu\lambda}$, $\tilde{B}^{\mu}:= g_{\alpha\beta}B^{\mu\alpha\beta}$.
\newline \\
\textbf{Proof:} Varying $(\ref{senar})$ with respect to the affine connection, we derive
\begin{equation}
P_{\lambda}^{\;\;\;\mu\nu}+B_{\lambda}^{\;\;\;\mu\nu}=0 \label{pala}
\end{equation}
where
\begin{equation}
B_{\lambda}^{\;\;\;\mu\nu}:=  \frac{2 \kappa}{\sqrt{-g}}\frac{\delta  S_{1}}{\delta \Gamma^{\lambda}_{\;\;\;\mu\nu}}=\frac{2 \kappa}{\sqrt{-g}}\frac{\partial (\sqrt{-g}\mathcal{L}_{1})}{\partial \Gamma^{\lambda}_{\;\;\;\mu\nu}}
\end{equation}
and $P_{\lambda}^{\;\;\;\mu\nu}$ is the Palatini tensor which is defined by
\begin{gather}
P_{\lambda}^{\;\;\;\mu\nu}:=  \frac{1}{\sqrt{-g}} \frac{\delta  S_{EH}}{\delta \Gamma^{\lambda}_{\;\;\;\mu\nu}} =\frac{1}{\sqrt{-g}} \frac{\partial( \sqrt{-g} R)}{\partial \Gamma^{\lambda}_{\;\;\;\mu\nu}}= \nonumber \\
=-\frac{\nabla_{\lambda}(\sqrt{-g}g^{\mu\nu})}{\sqrt{-g}}+\frac{\nabla_{\sigma}(\sqrt{-g}g^{\mu\sigma})}{\sqrt{-g}}\delta_{\lambda}^{\nu}+2(g^{\mu\nu}S_{\lambda}-S^{\mu}\delta^{\nu}_{\lambda}+g^{\mu\sigma}S_{\sigma\lambda}^{\;\;\;\nu})
\end{gather}
as we have already seen. Now, as we show in the appendix, the latter can also be written in the form
\begin{equation}
P^{\alpha\mu\nu}=\left( \frac{Q^{\alpha}}{2}+2 S^{\alpha}\right) g^{\mu\nu}-(Q^{\alpha\mu\nu}+2 S^{\alpha\mu\nu})+\left( \tilde{Q}^{\mu}-\frac{Q^{\mu}}{2}-2 S^{\mu} \right)g^{\nu\alpha}
\end{equation}
With this at hand and recalling the connection decomposition in terms of the Riemannian part, non-metricity and torsion
\begin{equation}
\Gamma^{\lambda}_{\;\;\;\mu\nu}=\tilde{\Gamma}^{\lambda}_{\;\;\;\mu\nu}+\frac{1}{2}g^{\alpha\lambda}(Q_{\mu\nu\alpha}+Q_{\nu\alpha\mu}-Q_{\alpha\mu\nu}) -g^{\alpha\lambda}(S_{\alpha\mu\nu}+S_{\alpha\nu\mu}-S_{\mu\nu\alpha}) \label{affconection}
\end{equation}
we observe that the combination $(Q^{\alpha\mu\nu}+2 S^{\alpha\mu\nu})$ appears in both and can, therefore, be eliminated. Indeed, pairing up a bit the terms of the last equation, we may re-write it as
\begin{equation}
\Gamma^{\lambda}_{\;\;\;\mu\nu}=\tilde{\Gamma}^{\lambda}_{\;\;\;\mu\nu}+\frac{1}{2}g^{\alpha\lambda}\Big( (Q_{\mu\nu\alpha}+2 S_{\mu\nu\alpha})+(Q_{\nu\alpha\mu}+2 S_{\nu\alpha\mu})-(Q_{\alpha\mu\nu}+2 S_{\alpha\mu\nu}) \Big) \label{solk}
\end{equation}
where we have used the fact that $S_{\alpha\mu\nu}=-S_{\mu\alpha\nu}$. In addition, we observe that 
\begin{equation}
P_{\alpha\mu\nu}-P_{\nu\alpha\mu}-P_{\mu\nu\alpha}=A_{\alpha\mu\nu}-g_{\alpha\mu}\tilde{Q}_{\nu}+2 g_{\nu[\alpha}(\tilde{Q}_{\mu]}-Q_{\mu]}-4 S_{\mu]})
\end{equation}
where
\begin{equation}
A_{\mu\nu\alpha}=(Q_{\mu\nu\alpha}+2 S_{\mu\nu\alpha})+(Q_{\nu\alpha\mu}+2 S_{\nu\alpha\mu})-(Q_{\alpha\mu\nu}+ 2 S_{\alpha\mu\nu})
\end{equation}
Thus, substituting the above combination into $(\ref{solk})$ we obtain
\begin{equation}
\Gamma^{\lambda}_{\;\;\;\mu\nu}=\tilde{\Gamma}^{\lambda}_{\;\;\;\mu\nu}+\frac{g^{\lambda\alpha}}{2}(P_{\alpha\mu\nu}-P_{\nu\alpha\mu}-P_{\mu\nu\alpha})+g^{\alpha\lambda}g_{\nu[\mu}(\tilde{Q}_{\alpha]}-Q_{\alpha]}-4 S_{\alpha]})+\frac{1}{2}\delta_{\mu}^{\lambda}\tilde{Q}_{\mu}
\end{equation}
Now, as we also prove in the appendix, it holds that
\begin{equation}
P^{\mu}\equiv P_{\lambda}^{\;\;\;\mu\lambda}=(n-1)\left( \tilde{Q}^{\mu}-\frac{1}{2}Q^{\mu}\right) +2(2-n)S^{\mu}
\end{equation}
\begin{equation}
\tilde{P}^{\mu}\equiv g_{\alpha\beta}P^{\mu\alpha\beta}=\frac{(n-3)}{2}Q^{\mu}+\tilde{Q}^{\mu}+2(n-2)S^{\mu}
\end{equation}
such that
\begin{equation}
P^{\mu}-\tilde{P}^{\mu}=(n-2)(\tilde{Q}^{\mu}-Q^{\mu}-4 S^{\mu})
\end{equation}
Using this fact, the connection recasts to
\begin{equation}
\Gamma^{\lambda}_{\;\;\;\mu\nu}=\tilde{\Gamma}^{\lambda}_{\;\;\;\mu\nu}+\frac{g^{\lambda\alpha}}{2}(P_{\alpha\mu\nu}-P_{\nu\alpha\mu}-P_{\mu\nu\alpha})+\frac{g^{\alpha\lambda}}{(n-2)}g_{\nu[\mu}(P_{\alpha]}-\tilde{P}_{\alpha]})+\frac{1}{2}\delta_{\mu}^{\lambda}\tilde{Q}_{\nu}  \label{gg}
\end{equation}
Notice now that our total action is projective invariant by assumption. This means, as we have already seen, that the theory is invariant under
\begin{equation}
\Gamma^{\lambda}_{\;\;\;\mu\nu} \rightarrow \Gamma^{\lambda}_{\;\;\;\mu\nu} +\delta^{\lambda}_{\mu}\xi_{\nu}
\end{equation}
for any vector $\xi_{\nu}$. That is, there exists an unspecified vectorial degree of freedom. Using this very fact we can always make any gauge choice that we may like. As it is apparent from ($\ref{gg}$) in order to get rid of the last term (which is unspecified) we make the gauge choice
\begin{equation}
\xi_{\nu}=-\frac{1}{2}\delta_{\mu}^{\lambda}\tilde{Q}_{\nu} 
\end{equation}
Then, the connection assumes the form
\begin{equation}
\Gamma^{\lambda}_{\;\;\;\mu\nu}=\tilde{\Gamma}^{\lambda}_{\;\;\;\mu\nu}+\frac{g^{\lambda\alpha}}{2}(P_{\alpha\mu\nu}-P_{\nu\alpha\mu}-P_{\mu\nu\alpha})+\frac{g^{\alpha\lambda}}{(n-2)}g_{\nu[\mu}(P_{\alpha]}-\tilde{P}_{\alpha]})
\end{equation}
Upon using ($\ref{pala}$) and defining $B^{\mu}:= B_{\lambda}^{\;\;\;\mu\lambda}$ along with
$\tilde{B}^{\mu}:= g_{\alpha\beta}B^{\mu\alpha\beta}$, we finally arrive at
\begin{equation}
\Gamma^{\lambda}_{\;\;\;\mu\nu}=\tilde{\Gamma}^{\lambda}_{\;\;\;\mu\nu}-\frac{g^{\lambda\alpha}}{2}(B_{\alpha\mu\nu}-B_{\nu\alpha\mu}-B_{\mu\nu\alpha})-\frac{g^{\alpha\lambda}}{(n-2)}g_{\nu[\mu}(B_{\alpha]}-\tilde{B}_{\alpha]}) \label{theo1}
\end{equation}
as stated.

\textbf{Comment 1:} The projective invariance of $S_{1}$ is only necessary in order to remove the term 
$\frac{1}{2}\delta_{\mu}^{\lambda}\tilde{Q}_{\nu}$ from $(\ref{gg})$. If  $S_{1}$  does not respect projective invariance one has to add the aforementioned term in the general result $(\ref{theo1})$.

\textbf{Comment 2:} If there is no gravitational sector to $S_{1}$, i.e the latter is a purely matter action $S_{1}=S_{M}$ then $B_{\lambda}^{\;\;\;\mu\nu}=-\kappa \Delta_{\lambda}^{\;\;\;\mu\nu}$ and the connection is found to be
\beq
\Gamma^{\lambda}_{\;\;\;\mu\nu}=\tilde{\Gamma}^{\lambda}_{\;\;\;\mu\nu}+\kappa\frac{g^{\lambda\alpha}}{2}(\Delta_{\alpha\mu\nu}-\Delta_{\nu\alpha\mu}-\Delta_{\mu\nu\alpha})+\frac{g^{\alpha\lambda}}{(n-2)}g_{\nu[\mu}(\Delta_{\alpha]}-\tilde{\Delta}_{\alpha]})
\eeq
where  $\Delta^{\mu}:= \Delta_{\lambda}^{\;\;\;\mu\lambda}$, $\tilde{\Delta}^{\mu}:= g_{\alpha\beta}\Delta^{\mu\alpha\beta}$ which, as it stands, is an algebraic equation for the connection given the fact that for a matter sector linear in $\Gamma$ the hypermomentum is independent of the connection.

\subsubsection{Expressions for torsion and non-metricity}
Having the above decomposition we can easily derive the expressions for torsion and non-metricity by their very definitions. Starting with torsion, we have
\begin{equation}
S_{\mu\nu}^{\;\;\;\;\lambda}:= \Gamma^{\lambda}_{\;\;\;[\mu\nu]}=\frac{1}{2}\Big( B_{[\mu\nu]}^{\;\;\;\;\;\lambda}+B_{[\nu \;\;\;\;\mu]}^{\;\;\;\lambda}-B^{\lambda}_{\;\;\;[\mu\nu]}\Big)-\frac{1}{2(n-2)}\delta^{\lambda}_{\nu}(B_{\mu}-\tilde{B}_{\mu})
\end{equation}
As long as non-metricity is concerned, from its definition it follows that
\begin{equation}
Q_{\alpha\mu\nu}:= -\nabla_{\alpha}g_{\mu\nu}=-\tilde{\nabla}_{\alpha}g_{\mu\nu}+\frac{1}{2}\Big( B_{\alpha\mu\nu}+B_{\nu\alpha\mu}+B_{\alpha\nu\mu}+B_{\mu\alpha\nu}-B_{\mu\nu\alpha}-B_{\nu\mu\alpha} \Big)
\end{equation}
Now, using the fact that the Levi-Civita connection is metric compatible ($\tilde{\nabla}_{\alpha}g_{\mu\nu}=0$) we obtain for the non-metricity
\begin{equation}
Q_{\alpha\mu\nu}=B_{(\mu\nu)\alpha}+B_{(\mu|\alpha|\nu)}-B_{\alpha(\mu\nu)}
\end{equation}

\textbf{Comment:} Since $S_{1}[g,\Gamma]$ is linear in the connection, its variation $B_{\alpha\mu\nu}$ is independent of the connection. Then, expression ($\ref{theo1}$) is an algebraic equation for the connection. So in this case, not surprisingly, the connection caries no dynamics.

\subsection{Generalizing the Theorem}
Now, our above result may be readily generalized for actions of the form
\begin{equation}
S[ g_{\mu\nu},\Gamma^{\lambda}_{\;\;\;\alpha\beta},\phi]=\frac{1}{2\kappa}\int d^{n}x\sqrt{-g}f(R)+S_{1}[ g_{\mu\nu},\Gamma^{\lambda}_{\;\;\;\alpha\beta},\phi] \label{sena}
\end{equation}
where we have replaced $R$ with a general $f(R)$ function. In addition, we will now consider the additional part $S_{1}$ to be a purely matter part, that is  $S_{1}[g,\Gamma,\phi]=S_{M}[g,\Gamma,\phi]$. We do so in order to see how the energy tensors (energy momentum and hyper-momentum) enter the picture, especially with regards to the dynamical content of the connection\footnote{Similar results hold if we consider also a gravitational sector to $S_{1}$ but then there is no direct contact with the energy tensors.}. So, we may now state and prove a second theorem.
\newline \\
\textbf{Theorem 2:} Consider the action
\begin{equation}
S[ g_{\mu\nu},\Gamma^{\lambda}_{\;\;\;\alpha\beta},\phi]=\frac{1}{2\kappa}\int d^{n}x\sqrt{-g}f(R)+S_{1}[ g_{\mu\nu},\Gamma^{\lambda}_{\;\;\;\alpha\beta},\phi] \label{sena2}
\end{equation}
where $\phi$ denotes any other additional fields that may be present in the spacetime and 
\begin{equation}
S_{1}[ g_{\mu\nu},\Gamma^{\lambda}_{\;\;\;\alpha\beta},\phi]=S_{M}[ g_{\mu\nu},\Gamma^{\lambda}_{\;\;\;\alpha\beta},\phi]=\frac{1}{2\kappa} \int d^{n}x\sqrt{-g} \mathcal{L}_{M}(g,\Gamma,\phi)
\end{equation}
Now given any general matter action $S_{1}[g,\Gamma,\phi]=S_{M}[g,\Gamma,\phi]$ that is
\begin{itemize}
	\item At most linear in $\Gamma^{\lambda}_{\;\;\;\mu\nu}$ and its partial derivatives
	\item Projective invariant
\end{itemize}
we state that the affine connection can solely be expressed in terms of $f^{'}(T)$ (where T is the trace of the energy momentum tensor) and of $\Gamma$-variations of $\mathcal{L}_{M}$ (i.e. the hypermomentum $\Delta_{\lambda}^{\;\;\;\mu\nu}$)  and its form is the following
\begin{equation}
\Gamma^{\lambda}_{\;\;\;\mu\nu}=\tilde{\Gamma}^{\lambda}_{\;\;\;\mu\nu}+\frac{g^{\lambda\alpha}}{2}(H_{\alpha\mu\nu}-H_{\nu\alpha\mu}-H_{\mu\nu\alpha})+\frac{g^{\alpha\lambda}}{(n-2)}g_{\nu[\mu}(H_{\alpha]}-\tilde{H}_{\alpha]})
\end{equation}
where
\begin{equation}
H_{\lambda}^{\;\;\;\mu\nu}:= - \frac{ 2 \kappa}{f^{'}\sqrt{-g}}\frac{\delta  \mathcal{L}_{M}}{\delta \Gamma^{\lambda}_{\;\;\;\mu\nu}}+\frac{1}{f^{'}}( g^{\mu\nu} \partial_{\lambda}f^{'}-\delta_{\lambda}^{\nu}\partial^{\mu} f^{'} )=\frac{\kappa}{f^{'}}\Delta_{\lambda}^{\;\;\;\mu\nu}+\frac{1}{f^{'}}( g^{\mu\nu} \partial_{\lambda}f^{'}-\delta_{\lambda}^{\nu}\partial^{\mu} f^{'})
\end{equation}
and $H^{\mu}:= H_{\lambda}^{\;\;\;\mu\lambda}$, \;$\tilde{H}^{\mu}:= g_{\alpha\beta}H^{\mu\alpha\beta}$,\; $f'=f'(T)$,\; $T:= g^{\mu\nu}T_{\mu\nu}=g^{\mu\nu}\frac{2}{\sqrt{-g}}\frac{\partial(\sqrt{-g}\mathcal{L}_{M})}{\partial g^{\mu\nu}}$  and the prime denotes differentiation with respect to the Ricci scalar.
\newline \\
\textbf{Proof:} Varying ($\ref{sena2}$) with respect to the connection we obtain
\beq
P_{\lambda}^{\;\;\;\mu\nu}(h)=\kappa \Delta_{\lambda}^{\;\;\;\mu\nu} \label{pafr}
\eeq
where
\begin{gather}
P_{\lambda}^{\;\;\;\mu\nu}(h) := -\frac{\nabla_{\lambda}(\sqrt{-g}f^{'}g^{\mu\nu})}{\sqrt{-g}}+\frac{\nabla_{\alpha}(\sqrt{-g}f^{'}g^{\mu\alpha}\delta_{\lambda}^{\nu})}{\sqrt{-g}}+ \\ \nonumber
2 f^{'}(S_{\lambda}g^{\mu\nu}-S^{\mu}\delta_{\lambda}^{\nu}-  S_{\lambda}^{\;\;\;\mu\nu}) \label{pakah1}
\end{gather}
 is the Palatini tensor of the metric $h_{\mu\nu}=f'(R)g_{\mu\nu}$, which is conformally related to $g_{\mu\nu}$ and prime here denotes differentiation with respect to the Ricci scalar. $ \Delta_{\lambda}^{\;\;\;\mu\nu}$ is the usual hypermomentum tensor we have defined earlier. Now, expanding the covariant derivatives in the above we see that
\beq
P_{\lambda}^{\;\;\;\mu\nu}(h)=f^{'}P_{\lambda}^{\;\;\;\mu\nu}(g)+\delta_{\lambda}^{\nu}g^{\mu\alpha}\partial_{\alpha}f^{'}-g^{\mu\nu}\partial_{\lambda}f^{'}
\eeq
where $P_{\lambda}^{\;\;\;\mu\nu}(g)$ is the usual Palatini tensor of $g_{\mu\nu}$. Then for $f'(R)\neq 0$ we may solve for the latter
\beq
P_{\lambda}^{\;\;\;\mu\nu}(g)=\frac{1}{f'}\left( P_{\lambda}^{\;\;\;\mu\nu}(h)-\delta_{\lambda}^{\nu}g^{\mu\alpha}\partial_{\alpha}f^{'}+g^{\mu\nu}\partial_{\lambda}f^{'}\right)
\eeq
or by virtue of ($\ref{pafr}$)
\beq
P_{\lambda}^{\;\;\;\mu\nu}(g)=\frac{1}{f'}\left( \kappa\Delta_{\lambda}^{\;\;\;\mu\nu}-\delta_{\lambda}^{\nu}g^{\mu\alpha}\partial_{\alpha}f^{'}+g^{\mu\nu}\partial_{\lambda}f^{'}\right)
\eeq
Then recalling eq. ($\ref{gg}$) that we obtained in the first Theorem,
\begin{equation}
\Gamma^{\lambda}_{\;\;\;\mu\nu}=\tilde{\Gamma}^{\lambda}_{\;\;\;\mu\nu}+\frac{g^{\lambda\alpha}}{2}\Big(P_{\alpha\mu\nu}(g)-P_{\nu\alpha\mu}(g)-P_{\mu\nu\alpha}(g)\Big)+\frac{g^{\alpha\lambda}}{(n-2)}g_{\nu[\mu}\Big(P_{\alpha]}(g)-\tilde{P}_{\alpha]}(g)\Big)+\frac{1}{2}\delta_{\mu}^{\lambda}\tilde{Q}_{\nu}  
\end{equation}  
and using the above, we find
\begin{gather}
\Gamma^{\lambda}_{\;\;\;\mu\nu}=\tilde{\Gamma}^{\lambda}_{\;\;\;\mu\nu}+\frac{\kappa}{f'}\frac{g^{\lambda\alpha}}{2}(\Delta_{\alpha\mu\nu}-\Delta_{\nu\alpha\mu}-\Delta_{\mu\nu\alpha})+\frac{\kappa}{f'}\frac{g^{\alpha\lambda}}{(n-2)}g_{\nu[\mu}(\Delta_{\alpha]}-\tilde{\Delta}_{\alpha]}) \nonumber \\
+\frac{1}{(n-2)f'}\Big( \delta^{\lambda}_{\nu}\partial_{\mu}f' -g_{\mu\nu} \partial^{\lambda}f' \Big)
+\frac{1}{2}\delta_{\mu}^{\lambda}\tilde{Q}_{\nu}  
\end{gather}  
where at this point $f'=f'(R)$. Now, variation of our total action with respect to the metric, yields
\beq
f^{'}(R)R_{(\mu\nu)}-\frac{f(R)}{2}g_{\mu\nu}=\kappa T_{\mu\nu}
\eeq
where 
\beq
T_{\mu\nu}:= -\frac{2}{\sqrt{-g}}\frac{\delta S_{M}}{\delta g^{\mu\nu}}
\eeq
which we may contract with the metric tensor to obtain
\beq
f^{'}(R)R-\frac{n}{2}f(R)=\kappa T \label{tmn}
\eeq
The latter defines the implicit function $R=R(T)$\footnote{Except in the case $f(R)\propto R^{2}$ for which the left send hide of ($\ref{tmn}$) is identically zero and the model allows only for conformally invariant matter ($T=0$). This exception have been studied in \cite{iosifidis2018torsion} where also the  cosmological solutions were given for this case.} and therefore  both $f(R)$ and $f^{'}(R)$  are all functions of $T$  ($f(R)=f(R(T))=f(T)$ and $f^{'}(R)=f^{'}(R(T))=f^{'}(T)$). With this at hand, and using the fact that our total action is projective invariant we may remove the term $\frac{1}{2}\delta_{\mu}^{\lambda}\tilde{Q}_{\nu} $ and write
\begin{gather}
\Gamma^{\lambda}_{\;\;\;\mu\nu}=\tilde{\Gamma}^{\lambda}_{\;\;\;\mu\nu}+\frac{\kappa}{f'}\frac{g^{\lambda\alpha}}{2}(\Delta_{\alpha\mu\nu}-\Delta_{\nu\alpha\mu}-\Delta_{\mu\nu\alpha})+\frac{\kappa}{f'}\frac{g^{\alpha\lambda}}{(n-2)}g_{\nu[\mu}(\Delta_{\alpha]}-\tilde{\Delta}_{\alpha]}) \nonumber \\
+\frac{1}{(n-2)f'}\Big( \delta^{\lambda}_{\nu}\partial_{\mu}f' -g_{\mu\nu} \partial^{\lambda}f' \Big) 
\end{gather} 
where $f'$ is a function of $T$ now. Finally, defining
\beq
H_{\alpha\mu\nu}:=\frac{1}{f'}\left( \kappa \Delta_{\alpha\mu\nu}+ g_{\mu\nu}\partial_{\alpha}f' -g_{\nu\alpha}\partial_{\mu}f' \right)
\eeq
we complete the proof
\begin{equation}
\Gamma^{\lambda}_{\;\;\;\mu\nu}=\tilde{\Gamma}^{\lambda}_{\;\;\;\mu\nu}+\frac{g^{\lambda\alpha}}{2}(H_{\alpha\mu\nu}-H_{\nu\alpha\mu}-H_{\mu\nu\alpha})+\frac{g^{\alpha\lambda}}{(n-2)}g_{\nu[\mu}(H_{\alpha]}-\tilde{H}_{\alpha]})
\end{equation}

\subsection{Generalized Theorem}
We may now relax our assumptions and let $S_{1}[ g_{\mu\nu},\Gamma^{\lambda}_{\;\;\;\alpha\beta},\phi]$ have an arbitrary dependence on the connection and its derivatives and may not respect the projective symmetry in general. This leads us to the third Theorem.
\newline \\
\textbf{Theorem 3:} Consider the action
\begin{equation}
S[ g_{\mu \nu},\Gamma^{\lambda}_{\;\;\;\alpha\beta},\phi]=\frac{1}{2\kappa}\int d^{n}x\sqrt{-g}R+S_{1}[ g_{\mu\nu},\Gamma^{\lambda}_{\;\;\;\alpha\beta},\phi] \label{sena}
\end{equation}
where
\begin{equation}
S_{1}[ g_{\mu\nu},\Gamma^{\lambda}_{\;\;\;\alpha\beta},\phi]=\frac{1}{2\kappa} \int d^{n}x\sqrt{-g} \mathcal{L}_{1}(g,\Gamma,\phi)
\end{equation}
has an arbitrary dependence on the affine connection and its derivatives. Then, the connection is given by the solution of
\begin{equation}
\Gamma^{\lambda}_{\;\;\;\mu\nu}=\tilde{\Gamma}^{\lambda}_{\;\;\;\mu\nu}-\frac{g^{\lambda\alpha}}{2}(B_{\alpha\mu\nu}-B_{\nu\alpha\mu}-B_{\mu\nu\alpha})-\frac{g^{\alpha\lambda}}{(n-2)}g_{\nu[\mu}(B_{\alpha]}-\tilde{B}_{\alpha]}) 
\end{equation}
where
\begin{equation}
B_{\lambda}^{\;\;\;\mu\nu}:=  \frac{2 \kappa}{\sqrt{-g}}\frac{\delta  S_{1}}{\delta \Gamma^{\lambda}_{\;\;\;\mu\nu}}=\frac{2 \kappa}{\sqrt{-g}}\frac{\partial (\sqrt{-g}\mathcal{L}_{1})}{\partial \Gamma^{\lambda}_{\;\;\;\mu\nu}}
\end{equation}
 $B^{\mu}:= B_{\lambda}^{\;\;\;\mu\lambda}$, $\tilde{B}^{\mu}:= g_{\alpha\beta}B^{\mu\alpha\beta}$
and the above will be a differential equation for the connection in general since B has an arbitrary dependence  on the connection and its derivatives. \newline \\
\textbf{Proof:} Following identical steps with  Theorem-1 but now keeping in mind that $B_{\lambda}^{\;\;\;\mu\nu}(\Gamma, \partial \Gamma)$ is a general function of the connection and its derivatives, we get
\begin{equation}
\Gamma^{\lambda}_{\;\;\;\mu\nu}=\tilde{\Gamma}^{\lambda}_{\;\;\;\mu\nu}-\frac{g^{\lambda\alpha}}{2}(B_{\alpha\mu\nu}-B_{\nu\alpha\mu}-B_{\mu\nu\alpha})-\frac{g^{\alpha\lambda}}{(n-2)}g_{\nu[\mu}(B_{\alpha]}-\tilde{B}_{\alpha]}) \label{theo3}
\end{equation}
where $B^{\mu}:= B_{\lambda}^{\;\;\;\mu\lambda}$,
$\;\tilde{B}^{\mu}:= g_{\alpha\beta}B^{\mu\alpha\beta}$ and since $B_{\lambda}^{\;\;\;\mu\nu}(\Gamma, \partial \Gamma)$ has an arbitrary dependence of the connection and its derivatives, the above is a dynamical equation for the connection in contrast to equation ($\ref{theo1}$) which is an algebraic one. Having presented and proved the three Theorems we may now see some examples where the latter can by applied.

\section{Example 1: Exciting Torsional d.o.f.}
Let us now use the results we obtained for the connection decomposition (the $3$ Theorems) in order to review the model studied in \cite{d1982gravity,leigh2009torsion,petkou2010torsional} but now in the coordinate formalism. It is easy to show that the Nieh-Yan term considered there, translates to
\begin{equation}
\epsilon^{\mu\nu\rho\sigma}\partial_{\mu}S_{\nu\rho\sigma}
\end{equation}
in the coordinate formalism. Therefore, in our formalism the total action reads
\begin{gather}
S=\frac{1}{2\kappa}\int d^{4}x \sqrt{-g}R_{(\mu\nu)}g^{\mu\nu}+\frac{1}{2\kappa}\int d^{4}x F(x)\epsilon^{\mu\nu\rho\sigma}\partial_{\mu}S_{\nu\rho\sigma}= \nonumber \\
=\frac{1}{2\kappa}\int d^{4}x \sqrt{-g}R_{(\mu\nu)}g^{\mu\nu}-\frac{1}{2\kappa}\int d^{4}x \epsilon^{\mu\nu\rho\sigma}(\partial_{\mu}F)S_{\nu\rho\sigma}+s.t.
\end{gather}
where $F(x)$ is a scalar and $s.t.$ stands for surface term. Notice that the additional piece here is linear in the connection and therefore falls in the category of our Theorem-1. So, we may proceed and use the result we obtained for the connection.
 Variation with respect to the connection yields\footnote{Where we have used the properties of the Levi-Civita symbol and also raised an index with the metric.}
\begin{equation}
\sqrt{-g}P^{\nu\rho\sigma}-\epsilon^{\alpha\rho\sigma\nu}(\partial_{\alpha}F)=0 \label{epsif}
\end{equation}
Also, since in this model the non-metricity is zero, the Palatini tensor reads
\begin{equation}
P^{\nu\rho\sigma}=2\Big( g^{\rho\sigma}S^{\nu}-g^{\sigma\nu}S^{\rho}+ S^{\rho\nu\sigma} \Big) \label{ddf}
\end{equation}
Now, contracting ($\ref{epsif}$) with $\epsilon^{\mu\rho\sigma\nu}$ and also using the above, we obtain
\begin{gather}
\varepsilon_{\mu\nu\rho\sigma}S^{\rho\sigma\nu}=3(\partial_{\mu}F)
\end{gather}
where $\varepsilon_{\mu\nu\rho\sigma}\equiv \sqrt{-g} \epsilon_{\mu\rho\sigma\nu}$ is the Levi-Civita tensor. So, we may also write
\begin{equation}
\epsilon^{\mu\nu\rho\sigma}S_{\nu\rho\sigma}=3\sqrt{-g}g^{\mu\nu}(\partial_{\nu}F)
\end{equation}
Substituting the latter in our action we arrive at
\begin{equation}
S=\frac{1}{2\kappa}\int d^{4}x\Big[ \sqrt{-g}R-\sqrt{-g}3 g^{\mu\nu}(\partial_{\mu}F)(\partial_{\nu}F) \Big]
\end{equation}
We can also immediately see that
\begin{equation}
S_{\mu}=0\;,\;\;P^{\mu\nu\alpha}=-\varepsilon^{\rho\mu\nu\alpha}\partial_{\rho}F \;,\;\;S_{\mu\nu\alpha}=-\frac{1}{2}\varepsilon_{\mu\nu\alpha\lambda}\partial^{\lambda}F
\end{equation}
which when plugged into the connection decomposition ($\ref{theo1}$) of Theorem-1 yield
\begin{equation}
\Gamma^{\lambda}_{\;\;\;\;\mu\nu}=\tilde{\Gamma}^{\lambda}_{\;\;\;\;\mu\nu}+\frac{1}{2}\varepsilon_{\mu\nu}^{\;\;\;\;\;\rho\lambda}\partial_{\rho}F
\end{equation}
From which we conclude that this kind of torsion (being totally antisymmetric) has no effect on the autoparallels and the latter coincide with the geodesics. Now, we can fully decompose our original action to a Riemannian part plus an axion field. Indeed, to see this first recall the Ricci scalar decomposition
\begin{equation}
R=\tilde{R}+ \tilde{\nabla}_{\mu}( A^{\mu}-B^{\mu})+ B_{\mu}A^{\mu}-N_{\alpha\mu\nu}N^{\mu\nu\alpha}
\end{equation}
Note now that the second term is a surface term and can therefore be dropped when taken into the action integral. Regarding the other quantities appearing, we compute for our case
\begin{equation}
N_{\mu\nu\alpha}=-\frac{1}{2}\varepsilon_{\mu\nu\alpha\rho}\partial^{\rho}F
\end{equation}
\begin{equation}
A^{\mu}= N^{\mu}_{\;\;\;\nu\beta}g^{\nu\beta}=0, \;\;\; B^{\mu}=N^{\alpha\mu}_{\;\;\;\;\alpha}=0
\end{equation}
\begin{equation}
N_{\alpha\mu\nu}N^{\mu\nu\alpha}=-\frac{3}{2}\partial_{\mu}F\partial^{\mu}F
\end{equation}
so that, when substituted back to our action give
\begin{equation}
S=\frac{1}{2\kappa}\int d^{4}x\sqrt{-g}\Big[ \tilde{R}-\frac{3}{2} g^{\mu\nu}(\partial_{\mu}F)(\partial_{\nu}F) \Big]
\end{equation}
which is the action of Einstein gravity plus an axionic massless field. To recap, for this model, the affine connection takes the form
\begin{equation}
\Gamma^{\lambda}_{\;\;\;\;\mu\nu}=\tilde{\Gamma}^{\lambda}_{\;\;\;\;\mu\nu}+ N^{\lambda}_{\;\;\;\;\mu\nu}=\tilde{\Gamma}^{\lambda}_{\;\;\;\;\mu\nu}+\frac{1}{2}\varepsilon_{\mu\nu}^{\;\;\;\;\rho\lambda}\partial_{\rho}F
\end{equation}
Note now that this type of torsion (totally) antisymmetric has no effect on the autoparallels and the latter coincide with the geodesics. However, for general torsion (even with vanishing non-metricity) the two are not the same.

\section{Example 2: Metric Affine f(R) Theories}
Let us now apply the results of our connection decomposition and study some characteristics of Metric Affine f(R) theories (\cite{sotiriou2007metric,olmo2011palatini,vitagliano2010dynamics,olmo2009dynamical}) . Firstly we consider the vacuum theories and then we add matter. Then we review the ways that have been proposed in order to break the projective invariance and formulate another way to do so.

\subsection{Vacuum f(R) Theories}
Since we are in vacuum, our starting action will be
\beq
S=\frac{1}{2\kappa}\int d^{n}x \sqrt{-g}f(R) \label{ffrr}
\eeq
Varying with respect to the metric and using the principle of least action, we obtain
\beq
\delta_{g}S=\frac{1}{2\kappa}\int d^{n}x\sqrt{-g}\left[ f^{'}(R)R_{(\mu\nu)}-\frac{f(R)}{2}g_{\mu\nu} \right]=0 
\eeq
or
\beq
f^{'}(R)R_{(\mu\nu)}-\frac{f(R)}{2}g_{\mu\nu}=0 \label{efr}
\eeq
Now, using the fact that for a general tensor field (or density) $B^{\mu\nu}$ it holds that
\beq
B^{\mu\nu}\delta_{\Gamma}R_{\mu\nu}=\delta \Gamma^{\lambda}_{\;\;\;\mu\nu}\Big( -\nabla_{\lambda}B^{\mu\nu}+\nabla_{\alpha}(B^{\mu\alpha}\delta_{\lambda}^{\nu})-2 B^{\mu\alpha}S_{\lambda\alpha}^{\;\;\;\;\nu}  \Big)+ A
\eeq
where 
\beq
A=\nabla_{\lambda}( B^{\mu\nu}\delta \Gamma^{\lambda}_{\;\;\;\mu\nu}-B^{\mu\lambda}\delta_{\alpha}^{\nu}\delta \Gamma^{\alpha}_{\;\;\;\mu\nu})
\eeq
we vary with respect to $\Gamma^{\alpha}_{\;\;\;\mu\nu}$, to get
\beq
-\nabla_{\lambda}(\sqrt{-g}f^{'}g^{\mu\nu})+\nabla_{\alpha}(\sqrt{-g}f^{'}g^{\mu\alpha}\delta_{\lambda}^{\nu})+\\ \nonumber
2 \sqrt{-g}f^{'}(S_{\lambda}g^{\mu\nu}-S^{\mu}\delta_{\lambda}^{\nu}-  S_{\lambda}^{\;\;\;\mu\nu})=0 \label{eGf}
\eeq
Now we wish to solve the system of equations $(\ref{efr})$ and $(\ref{eGf})$. To do so, we first take the trace of $(\ref{efr})$ to arrive at
\beq
f^{'}(R)R-\frac{n}{2}f(R)=0 \label{rre}
\eeq
This is an algebraic equation on $R$ and it will have a number of solutions\footnote{When this equation has no solutions inconsistencies will arise as shown in \cite{ferraris1994universality}.} $R=R_{\kappa}=c_{\kappa}=constant$, \; $\kappa=1,2,...,i$ where $i$ is the number of solutions. Notice that for the specific choice $f(R) \propto R^{n/2}$ the above is identically satisfied. This case was studied extensively and the cosmological solutions were also given in \cite{iosifidis2018torsion}. So, going back to our solutions, for $R=R_{\kappa}=c_{\kappa}=constant$ and using the above equation, the field equations (\ref{efr}) take the form
\beq
R_{(\mu\nu)}-\frac{R_{\kappa}}{n}g_{\mu\nu}=0
\eeq
Also, since $f^{'}(R_{\kappa})$ is constant too, it can be pulled outside of the covariant derivative and ($\ref{eGf}$) becomes
\beq
f^{'}(R_{\kappa})\sqrt{-g} P_{\lambda}^{\;\;\;\mu\nu}=0
\eeq
where 
\beq
P_{\lambda}^{\;\;\;\mu\nu}=-\frac{\nabla_{\lambda}(\sqrt{-g}g^{\mu\nu})}{\sqrt{-g}}+\frac{\nabla_{\sigma}(\sqrt{-g}g^{\mu\sigma})\delta^{\nu}_{\lambda}}{\sqrt{-g}} \\
+2(S_{\lambda}g^{\mu\nu}-S^{\mu}\delta_{\lambda}^{\nu}+g^{\mu\sigma}S_{\sigma\lambda}^{\;\;\;\;\nu})
\eeq
is the Palatini tensor which we had defined earlier. This last equation implies
\beq
P_{\lambda}^{\;\;\;\mu\nu}=0
\eeq
which in turn, as we have shown, says that the geometry is Riemannian but with an undetermined vectorial degree of freedom. More specifically, as we showed in the previous section, the vanishing of the Palatini tensor implies that
\beq
R_{(\mu\nu)}=\tilde{R}_{\mu\nu}\;,\; R=\tilde{R}
\eeq
and our field equations reduce to
\beq
\tilde{R}_{\mu\nu}-\frac{c_{\kappa}}{n}g_{\mu\nu}=0
\eeq
The above field equations, are Einstein's field equations with a cosmological constant. In fact, this is GR with a whole set of Cosmological constants, for each solution $R=R_{\kappa}$ we pick we have a different theory with a Cosmological constant $\Lambda_{\kappa}=\frac{C_{\kappa}}{n}$. For a good discussion on this feature see also  
\cite{ferraris1994universality}. So, this is an interesting result especially when compared to metric $f(R)$ theories of Gravity in vacuum. In  metric $f(R)$ theories in vacuum the field equations are of forth order and of course they are different from Einstein equations. On the other hand, Metric-Affine $f(R)$ theories in vacuum, are equivalent  to a class of Einstein Gravities, with different Cosmological constants which are solutions of ($\ref{rre}$) and each solution gives a different value for the Cosmological constant. In fact, we have $i$-different theories, where $i$ is the number of solutions of ($\ref{rre}$). One important point take home though, is that in each of these there is an undetermined vectorial degree of freedom which does not interfere with Einstein equations at this point but nevertheless it is there, and will cause inconsistence theories when matter is added as we will see later.

\subsection{Metric Affine f(R) Theories With Matter}
Let us now try to add a matter term to the gravity action  ($\ref{ffrr}$) and derive the field equations for Metric Affine theories with matter. Note that this matter action can depend both on the metric tensor and the connection $S_{M}=S_{M}[g_{\alpha\beta},\Gamma^{\lambda}_{\;\;\;\mu\nu}]$ and its variation with respect to the metric tensor defines as usual the energy-momentum tensor while the variation with respect to the connection gives the hypermomentum tensor. So, our full action will be
\beq
S=S_{G}+S_{M}=\frac{1}{2\kappa}\int d^{n}x \sqrt{-g}f(R) +\int d^{n}x \sqrt{-g} \mathcal{L}_{M}
\eeq
Varying the above with respect to the metric tensor, we obtain
\beq
f^{'}(R)R_{(\mu\nu)}-\frac{f(R)}{2}g_{\mu\nu}=\kappa T_{\mu\nu}
\eeq
where 
\beq
T_{\mu\nu} := -\frac{2}{\sqrt{-g}}\frac{\delta S_{M}}{\delta g^{\mu\nu}}
\eeq
the usual energy-momentum (or stress-energy) tensor. Variation with respect to the independent connection gives
\beq
-\frac{\nabla_{\lambda}(\sqrt{-g}f^{'}g^{\mu\nu})}{\sqrt{-g}}+\frac{\nabla_{\alpha}(\sqrt{-g}f^{'}g^{\mu\alpha}\delta_{\lambda}^{\nu})}{\sqrt{-g}}+\\ \nonumber
2 f^{'}(S_{\lambda}g^{\mu\nu}-S^{\mu}\delta_{\lambda}^{\nu}-  S_{\lambda}^{\;\;\;\mu\nu})=\kappa \Delta_{\lambda}^{\;\;\;\mu\nu} \label{frmt}
\eeq
where
\beq
\Delta_{\lambda}^{\;\;\;\mu\nu} \equiv -\frac{2}{\sqrt{-g}}\frac{\delta S_{M}}{\delta \Gamma^{\lambda}_{\;\;\;\mu\nu}}
\eeq
is the hypermomentum tensor which gives information about the spin, shear and dilation currents of matter. Notice now that the left hand side of ($\ref{frmt}$) is the Palatini tensor computed for the modified tensor\footnote{This is just a mathematical convenience, $h_{\mu\nu}$ has no physical significance.}
\beq
h_{\mu\nu}=f^{'}(R)g_{\mu\nu}
\eeq
With this observation, we may write
\beq
P_{\lambda}^{\;\;\;\mu\nu}(h)=\kappa\Delta_{\lambda}^{\;\;\;\mu\nu} 
\eeq
where 
\begin{gather}
P_{\lambda}^{\;\;\;\mu\nu}(h) := -\frac{\nabla_{\lambda}(\sqrt{-g}f^{'}g^{\mu\nu})}{\sqrt{-g}}+\frac{\nabla_{\alpha}(\sqrt{-g}f^{'}g^{\mu\alpha}\delta_{\lambda}^{\nu})}{\sqrt{-g}}+ \\ \nonumber
2 f^{'}(S_{\lambda}g^{\mu\nu}-S^{\mu}\delta_{\lambda}^{\nu}-  S_{\lambda}^{\;\;\;\mu\nu}) 
\end{gather}
and by applying the product rule for the covariant derivatives we find
\beq
P_{\lambda}^{\;\;\;\mu\nu}(h)=f^{'}P_{\lambda}^{\;\;\;\mu\nu}(g)+\delta_{\lambda}^{\nu}g^{\mu\alpha}\partial_{\alpha}f^{'}-g^{\mu\nu}\partial_{\lambda}f^{'}
\eeq
where $P_{\lambda}^{\;\;\;\mu\nu}(g)$ is the usual Palatini tensor computed with respect to the metric tensor $g_{\mu\nu}$.
Now, as we have already  seen the Palatini tensor has zero trace when contracted in its two fist indices, that is\footnote{This is true irrespective of the metric used since $g_{\mu\nu}$ and $h_{\mu\nu}$ are conformally related.}
\beq
P_{\mu}^{\;\;\;\mu\nu}=0
\eeq
this is so because of the projective invariance of  the Ricci scalar $R$, and the above holds as an identity. This enforces
\beq
\Delta_{\mu}^{\;\;\;\mu\nu} =0
\eeq
and this, obviously, cannot be correct for any form of matter. We can find many examples of matter for which $\Delta_{\mu}^{\;\;\;\mu\nu} \neq 0$. For instance, suppose that we have a vector field $A_{\mu}$ whose matter action contains a term that goes like
\beq
S_{M}[g_{\alpha\beta},\Gamma^{\lambda}_{\;\;\;\mu\nu}]=-\frac{1}{4}\int d^{n}x \sqrt{-g} g^{\mu\alpha}g^{\nu\beta}(\nabla_{\mu}A_{\nu})(\nabla_{\alpha}A_{\beta})
\eeq 
The associated hypermomentum in this case, will be
\beq
\Delta_{\lambda}^{\;\;\;\mu\nu}=A_{\lambda}g^{\mu\alpha}g^{\nu\beta}(\nabla_{\beta}A_{\alpha})
\eeq
and therefore
\beq
\Delta_{\mu}^{\;\;\;\mu\nu}=A^{\alpha}(\nabla_{\beta}A_{\alpha})g^{\beta\nu} \neq 0 \label{conseqw}
\eeq
So, we see that when one tries to add matter to Metric Affine $f(R)$ Gravities inconsistency\footnote{Inconsistency may be too strong a word here. As pointed out in \cite{jimenez2018teleparallel} these constraints on the matter fields, like eq.$(\ref{conseqw})$, are perfectly fine even desirable in some cases (see also \cite{jimenez2017born} for a similar discussion). In addition all standard matter fields, both bosonic and
	fermionic, respect the projective symmetry so no consistency
	problem arises. So, whether projective invariance should be broken or not is an interesting open subject. However, its discussion goes beyond the scope of this paper. In these notes we just present an another way to break the invariance given that one wants to break it.} arises due to the projective invariance of the Ricci scalar (and of course any function-$f(R)$ of it will respect this invariance too). To obtain a self-consistent theory one needs to somehow break this projective invariance by fixing a vectorial degree of freedom. This can be done by adding extra terms in the action that do not respect the projective invariance, but this is somewhat arbitrary. What seems more natural to do is to fix either the torsion  or non-metricity vectors to zero by means of a Lagrange multiplier added to the matter action. In \cite{1981GReGr..13.1037H,Lord2004MetricAftM} they fixed the Weyl vector $Q_{\mu}$ to zero\footnote{A similar way of breaking the projective invariance was also presented in \cite{smalley1979volume}.} but in \cite{sotiriou2007metric} it was shown that this is not a viable choice and works only for $f(R)=R$ that is, only for the Einstein Hilbert action, and the best way to proceed is to set $S_{\mu}=0$ by means of a Lagrange multiplier \cite{sotiriou2007metric}. We review both of them in the following chapter, along with some other possibility.

\subsection{Breaking the Projective Invariance}
In order to break the projective invariance one needs to fix a vectorial degree of freedom. So, what vectors do we have at our disposal? As we have seen, we can construct two vectors out of non-metricity by contracting with the metric. These are the Weyl
\beq
 Q_{\alpha}=Q_{\alpha\mu\nu}g^{\mu\nu}
\eeq
and the second non-metricity vector 
\beq
\tilde{Q}_{\nu}=Q_{\alpha\mu\nu}g^{\alpha\mu}
\eeq
For torsion, because of its antisymmetry there is simply one vector to be constructed by contractions, and this is the torsion vector
\beq
S_{\mu}=S_{\mu\lambda}^{\;\;\;\;\lambda}
\eeq
There is also another possibility (in 4-dim), by contracting the torsion tensor with the Levi-Civita symbol we get the pseudo-vector
\beq
\tilde{S}^{\alpha} =-\epsilon^{\mu\nu\lambda\alpha}S_{\mu\nu\lambda}
\eeq
However, this quantity is itself invariant under projective transformations of the connection and therefore it cannot be used to break the projective invariance. As a result, the vectors that could potentially break the projective invariance and produce a self-consistent theory, are $\{ Q_{\alpha},\tilde{Q}_{\nu},S_{\mu}\}$ . We explore the possibility of fixing each of them to zero separately.

\subsubsection{Fixing $S_{\mu}=0$}
Let us now break the projective invariance and obtain a self-consistent theory by fixing the torsion vector to zero, as done in \cite{sotiriou2007metric}. To this end we add the part
\beq
S_{B}=\int d^{n}x\sqrt{-g}B_{\mu}S^{\mu}
\eeq
where $B_{\mu}$ is a Lagrange multiplier that will fix $S_{\mu}$ to zero. Therefore, our total action will be
\begin{gather}
S[g_{\alpha\beta},\Gamma^{\lambda}_{\;\;\;\mu\nu},B_{\rho}]=S_{G}+S_{M}+S_{B}= \\ \nonumber
=\int  d^{n}x\sqrt{-g} \left[ \frac{1}{2\kappa}f(R)+\mathcal{L}_{M}+B_{\mu}S^{\mu}\right]
\end{gather}
and the total variation will have three different parts to it
\beq
\delta S=\delta_{g}S+\delta_{\Gamma}S+\delta_{B}S
\eeq
so the least action principle will give
\beq
\delta S=0 \Rightarrow \;\delta_{g}S=0\;,\;\delta_{\Gamma}S=0\;,\;\delta_{B}S=0
\eeq
Now, the parts $S_{G}$ and $S_{M}$ we have already varied in the previous chapter, so we only need to focus on the variation of $S_{B}$, which contains the parts
\beq
\delta S_{B}=\delta_{g} S_{B}+\delta_{\Gamma} S_{B}+\delta_{B} S_{B}
\eeq
and an easy calculation reveals
\beq
\delta_{g} S_{B}= \int d^{n}x\sqrt{-g}(\delta g^{\mu\nu} )\left[ -\frac{1}{2}g_{\mu\nu}B_{\alpha}S^{\alpha}+B_{(\mu}S_{\nu)} \right]
\eeq
\beq
\delta_{\Gamma} S_{B}= \int d^{n}x\sqrt{-g}(\delta \Gamma^{\lambda}_{\;\;\;\mu\nu}) \Big[ B^{[\mu}\delta^{\nu]}_{\lambda} \Big]
\eeq
and
\beq
\delta_{B} S_{B}= \int d^{n}x\sqrt{-g}(\delta B^{\mu})S_{\mu}
\eeq
respectively. So, varying the total action independently with respect to $g_{\alpha\beta},\;\Gamma^{\lambda}_{\;\;\;\mu\nu}$ and $B_{\rho}$ and applying the Least Action Principle, we obtain the set of field equations
\beq
f^{'}(R)R_{(\mu\nu)}-\frac{f(R)}{2}g_{\mu\nu}=\kappa \left( T_{\mu\nu}-\frac{1}{2}g_{\mu\nu}B_{\alpha}S^{\alpha}-B_{(\mu}S_{\nu)} \right)
\eeq 
\begin{gather}
-\frac{\nabla_{\lambda}(\sqrt{-g}f^{'}g^{\mu\nu})}{\sqrt{-g}}+\frac{\nabla_{\alpha}(\sqrt{-g}f^{'}g^{\mu\alpha}\delta_{\lambda}^{\nu})}{\sqrt{-g}}+
2 f^{'}(S_{\lambda}g^{\mu\nu}-S^{\mu}\delta_{\lambda}^{\nu}-  S_{\lambda}^{\;\;\;\mu\nu}) = \nonumber \\
 \kappa ( \Delta_{\lambda}^{\;\;\;\mu\nu}-  B^{[\mu}\delta^{\nu]}_{\lambda}  )
\end{gather}
\beq
 S_{\mu}=0
\eeq
Using the last equation ($S_{\mu}=0$)  the first two simplify and give
\beq
f^{'}(R)R_{(\mu\nu)}-\frac{f(R)}{2}g_{\mu\nu}=\kappa  T_{\mu\nu}
\eeq
\begin{gather}
-\frac{\nabla_{\lambda}(\sqrt{-g}f^{'}g^{\mu\nu})}{\sqrt{-g}}+\frac{\nabla_{\alpha}(\sqrt{-g}f^{'}g^{\mu\alpha}\delta_{\lambda}^{\nu})}{\sqrt{-g}}-
2 f^{'} S_{\lambda}^{\;\;\;\mu\nu} = \nonumber \\
 \kappa ( \Delta_{\lambda}^{\;\;\;\mu\nu}-  B^{[\mu}\delta^{\nu]}_{\lambda}  ) \label{Pam}
\end{gather} 
Now, taking the trace $\mu=\lambda$ in the last one, the left hand side is identically zero (since this is the contraction the modified Palatini tensor $P_{\mu}^{\;\;\;\mu\nu}(h)$) and we are left with
\beq
B^{\mu}=\frac{2}{1-n}\Delta_{\mu}^{\;\;\;\mu\nu}=\frac{2}{1-n}\tilde{\Delta}^{\nu}
\eeq
where we defined  $\Delta_{\mu}^{\;\;\;\mu\nu} := \tilde{\Delta}^{\nu}$. Thus, this is the value we should pick for the Lagrange multiplier $B_{\mu}$ in order to obtain self-consistent field equations, which upon this last substitution, take their final form
\beq
f^{'}(R)R_{(\mu\nu)}-\frac{f(R)}{2}g_{\mu\nu}=\kappa  T_{\mu\nu} \label{emde}
\eeq
\begin{gather}
-\frac{\nabla_{\lambda}(\sqrt{-g}f^{'}g^{\mu\nu})}{\sqrt{-g}}+\frac{\nabla_{\alpha}(\sqrt{-g}f^{'}g^{\mu\alpha}\delta_{\lambda}^{\nu})}{\sqrt{-g}}-
2 f^{'} S_{\lambda}^{\;\;\;\mu\nu} = \nonumber \\
 \kappa \Big( \Delta_{\lambda}^{\;\;\;\mu\nu}+  \frac{2}{n-1}\tilde{\Delta}^{[\mu}\delta^{\nu]}_{\lambda} \Big) \label{Pam}
\end{gather}  
Along with the constraint $S_{\mu}=0$ this is a set of consistent field equations, whose dynamics have studied to some extend in \cite{sotiriou2007metric,vitagliano2011dynamics}. We will review it here and add some new calculations regarding the form of non-metricity when the matter action does not depend on the connection. More specifically, we claim that when the connection is decoupled from the matter action ($\Delta_{\lambda}^{\;\;\;\mu\nu}=0$) torsion vanishes and the non-metricity is not general but we have  the case of a  Weyl non-metricity. To prove this, setting the right hand side of ($\ref{Pam}$) equal to zero , we obtain
\begin{gather}
-\frac{\nabla_{\lambda}(\sqrt{-g}f^{'}g^{\mu\nu})}{\sqrt{-g}}+\frac{\nabla_{\alpha}(\sqrt{-g}f^{'}g^{\mu\alpha}\delta_{\lambda}^{\nu})}{\sqrt{-g}}-
2 f^{'} S_{\lambda}^{\;\;\;\mu\nu} = 0 \label{Pamp}
\end{gather} 
and contracting in $\lambda=\nu$
\beq
\frac{(n-1)}{2}\frac{\nabla_{\alpha}(\sqrt{-g}f^{'}g^{\mu\alpha})}{\sqrt{-g}}-2 f^{'}S_{\lambda}^{\;\;\;\mu\lambda}=0
\eeq
but noticing that
\beq
S_{\lambda}^{\;\;\;\mu\lambda}=g_{\lambda\alpha}S^{\alpha\mu\lambda}=-g_{\lambda\alpha}S^{\mu\alpha\lambda}=-g^{\mu\kappa}S_{\kappa\lambda}^{\;\;\;\lambda} =-g^{\mu\kappa}S_{\kappa}=-S^{\mu}=0
\eeq
and substituting it above, we are left with
\beq
\frac{\nabla_{\alpha}(\sqrt{-g}f^{'}g^{\mu\alpha})}{\sqrt{-g}}=0
\eeq
which when itself is substituted back in ($\ref{Pamp}$) simplifies it to
\beq
\frac{\nabla_{\lambda}(\sqrt{-g}f^{'}g^{\mu\nu})}{\sqrt{-g}}+2 f^{'} S_{\lambda}^{\;\;\;\mu\nu} =0 \label{Plt}
\eeq
Taking the antisymmetric part in $\mu,\nu$ of the above we conclude that
\beq
 S_{\lambda\mu\nu}=S_{\lambda\nu\mu} 
\eeq
That is, torsion has to be symmetric on its second and third indices. But recall that torsion is antisymmetric when exchanging  first and second index. Any rank $3$ tensor that has both of these symmetries has to identically vanish. To see this, given that
\beq
S_{\mu\nu\lambda}=-S_{\nu\mu\lambda}\;,\; S_{\mu\nu\lambda}=S_{\mu\lambda\nu}
\eeq
exploiting these symmetries, we have
\begin{gather}
S_{\mu\nu\lambda}=S_{\mu\lambda\nu}=-S_{\lambda\mu\nu}=-S_{\lambda\nu\mu}=+S_{\nu\lambda\mu} = \nonumber \\
=S_{\nu\mu\lambda}=-S_{\mu\nu\lambda} 
\end{gather}
that is
\beq
S_{\mu\nu\lambda}=0
\eeq
Thus, torsion vanishes and ($\ref{Plt}$) becomes
\beq
\nabla_{\lambda}(\sqrt{-g}f^{'}g^{\mu\nu})=0
\eeq
This very condition tells us that the non-metricity has to be of the Weyl type ( namely $Q_{\alpha\mu\nu}\propto Q_{\alpha}g_{\mu\nu}$ ). To see this, expand the covariant derivative
\beq
g^{\mu\nu}f^{'}\nabla_{\lambda}\sqrt{-g}+Q_{\lambda}^{\;\;\;\;\mu\nu}+g^{\mu\nu}\partial_{\lambda}f^{'}=0
\eeq
and use
\beq
\frac{\nabla_{\lambda}\sqrt{-g}}{\sqrt{-g}}=-\frac{1}{2}Q_{\lambda}
\eeq
to arrive at
\beq
-\frac{1}{2}Q_{\lambda} g^{\mu\nu}+Q_{\lambda}^{\;\;\;\;\mu\nu}+g^{\mu\nu}\frac{\partial_{\lambda}f^{'}}{f^{'}}=0
\eeq
Contracting this with the metric tensor $g_{\mu\nu}$ it follows that
\beq
Q_{\lambda}=\frac{2n}{n-2}\partial_{\lambda}\ln{f^{'}}
\eeq
Finally, substituting the latter in the former we get
\beq
Q_{\lambda\mu\nu}=\frac{Q_{\lambda}}{n}g_{\mu\nu}=\frac{2}{n-2}g_{\mu\nu}\partial_{\lambda}\ln{f^{'}}
\eeq
In addition, contraction of ($\ref{emde}$) with the metric tensor gives
\beq
f^{'}(R)R-\frac{n}{2}f(R)=\kappa T
\eeq
which defines the implicit function $R=R(T)$ and therefore  both $f(R)$ and $f^{'}(R)$  are functions of $T$  ($f(R)=f(R(T))=f(T)$ and $f^{'}(R)=f^{'}(R(T))=f^{'}(T)$). As a result, a given $T_{\mu\nu}$ will give rise to Weyl non-metricity
\beq
Q_{\lambda\mu\nu}=\frac{Q_{\lambda}}{n}g_{\mu\nu}=\frac{2}{n-2}g_{\mu\nu}\partial_{\lambda}\ln{f^{'}(T)}
\eeq
In fact, this is a  Weyl Integrable Geometry (WIG) since the Weyl vector is exact ($Q_{\mu}\propto \partial_{\mu}\ln{f^{'}}$). So, to conclude, we have shown that a general $f(R)$ theory for which $S_{\mu}$ is fixed to zero and the matter fields do not couple to the connection ($\Delta_{\lambda}^{\;\;\;\mu\nu}=0$)  results in a theory with zero torsion and a Weyl Integrable Geometry. We now explore the possibility of fixing either of the two non-metricity vectors ($Q_{\mu}\;$, $\tilde{Q}_{\mu}$) to zero.

\subsubsection{Fixing $\tilde{Q}_{\mu}=0$ or $Q_{\mu}=0$}
We now add the Lagrange multiplier $C_{\mu}$ and the new piece to our action is
\beq
S_{C}=\int d^{n}\sqrt{-g}C_{\mu}\tilde{Q}^{\mu}
\eeq
We could may as well have replaced $\tilde{Q}^{\mu}$ with $Q_{\mu}$ (this was the fixing proposed in \cite{1981GReGr..13.1037H}) in the above but identical results will follow as we show below.  Again, let us consider the vacuum case where the Lagrange multiplier itself vanishes.\footnote{Not a-priori but after taking the trace and expressing it in terms of the Hypermomentum as we saw before.} Varying with respect to the connection and the Lagrange multiplier respectively we derive
\begin{gather}
-\frac{\nabla_{\lambda}(\sqrt{-g}f^{'}g^{\mu\nu})}{\sqrt{-g}}+\frac{\nabla_{\alpha}(\sqrt{-g}f^{'}g^{\mu\alpha}\delta_{\lambda}^{\nu})}{\sqrt{-g}}+
2 f^{'}(S_{\lambda}g^{\mu\nu}-S^{\mu}\delta_{\lambda}^{\nu}-  S_{\lambda}^{\;\;\;\mu\nu}) = 
 0 \label{Pampe}
\end{gather}
\beq
\tilde{Q}_{\mu}=0
\eeq
Now, even though we have set $\tilde{Q}_{\mu}=0$ we will keep $\tilde{Q}_{\mu}$  in our calculations to see what causes the problem when one tries to fix to zero either of the non-metricity vectors. To this end, contacting ($\ref{Pampe}$) in $\lambda=\nu$ we get 
\beq
\frac{\nabla_{\alpha}(\sqrt{-g}f^{'}g^{\mu\alpha})}{\sqrt{-g}}=2 f^{'}\frac{(n-2)}{n-1}S^{\mu} \label{recal}
\eeq
which when substituted back above, gives
\beq
-\frac{\nabla_{\lambda}(\sqrt{-g}f^{'}g^{\mu\nu})}{\sqrt{-g}}+2 f^{'}(S_{\lambda}g^{\mu\nu}+\frac{1}{1-n}S^{\mu}\delta_{\lambda}^{\nu}-  S_{\lambda}^{\;\;\;\mu\nu})=0
\eeq
After expanding the term in the covariant derivative and using the definitions of non-metricity, the above recasts to
\beq
\frac{1}{2}Q_{\lambda}g^{\mu\nu} -Q_{\lambda}^{\;\;\;\;\mu\nu}-g^{\mu\nu}\frac{\partial_{\lambda}f^{'}}{f^{'}}          +2(S_{\lambda}g^{\mu\nu}+\frac{1}{1-n}S^{\mu}\delta_{\lambda}^{\nu}-  S_{\lambda}^{\;\;\;\mu\nu})=0 \label{kku}
\eeq
where we have also divided through by $f^{'}$. Contracting the latter with the metric tensor $g^{\mu\nu}$ it follows that
\beq
\frac{(n-2)}{2}Q_{\lambda}-n\frac{\partial_{\lambda}f^{'}}{f^{'}}+\frac{2 n(n-2)}{(n-1)}S_{\lambda}=0 \label{eqw1}
\eeq
Also, contracting ($\ref{kku}$) in $\lambda=\nu$ we obtain
\beq
-\frac{1}{2}Q^{\mu}+\tilde{Q}^{\mu}+\frac{\partial^{\mu} f^{'}}{f^{'}}-\frac{2(n-2)}{(n-1)}S^{\mu}=0
\eeq
Multiplying through by $n$ and bringing the index downstairs, we may write the last one as
\beq
-\frac{n}{2}Q_{\lambda}+n \tilde{Q}_{\lambda}+n\frac{\partial_{\lambda} f^{'}}{f^{'}}-\frac{2 n(n-2)}{(n-1)}S_{\lambda}=0 \label{eqw2}
\eeq
Therefore, adding up equations ($\ref{eqw1}$) and ($\ref{eqw2}$) it follows that
\beq
-Q_{\lambda}+n \tilde{Q}_{\lambda}=0
\eeq
From this we see that fixing either of $Q_{\lambda}$ or $\tilde{Q}_{\lambda}$ to zero, the other vector must vanish too. So, by adding either of the Lagrange multipliers the end result is the same $\tilde{Q}_{\mu}=Q_{\mu}=0$, and with this at hand, from $(\ref{eqw2})$ we conclude that
\beq
\frac{\partial_{\mu}f^{'}}{f^{'}}=2\frac{(n-2)}{(n-1)}S_{\mu}
\eeq
Substituting all of these back into $(\ref{kku})$ it follows that
\beq
Q_{\lambda}^{\;\;\;\;\mu\nu}+2 S_{\lambda}^{\;\;\;\mu\nu}=\frac{2}{n-1}\Big[ S_{\lambda}g^{\mu\nu}-S^{\mu}\delta^{\nu}_{\lambda} \Big]
\eeq
or 
\beq
Q^{\alpha\mu\nu}+2 S^{\alpha\mu\nu}=\frac{2}{n-1}\Big[ S^{\alpha}g^{\mu\nu}-S^{\mu}g^{\alpha\nu} \Big] \label{qsw}
\eeq
Taking the symmetric part in $\alpha,\mu$ in the above we obtain
\beq
Q^{(\alpha\mu)\nu}=0
\eeq
where we have also used the fact that the torsion tensor is antisymmetric in its first two indices $(S_{(\alpha\mu)\nu}=0)$. The above equation implies that non-metricity has to be antisymmetric in its first two indices, but by definition it is symmetric in its last two. Any rank-$3$ tensor with such properties must identically vanish. Indeed, given that
\beq
Q_{\alpha\mu\nu}=-Q_{\mu\alpha\nu}\;\;and\;\; Q_{\alpha\mu\nu}=Q_{\alpha\nu\mu}
\eeq
we compute
\beq
Q_{\alpha\mu\nu}=-Q_{\mu\alpha\nu}=-Q_{\mu\nu\alpha}=Q_{\nu\mu\alpha}=Q_{\nu\alpha\mu}=-Q_{\alpha\nu\mu}=-Q_{\alpha\mu\nu}
\eeq
and therefore
\beq
Q_{\alpha\mu\nu}=0
\eeq
and we see that the whole non-metricity vanishes. In addition, taking the antisymmetric part of ($\ref{Pampe}$) and contracting in $\lambda=\mu$ we have 
\beq
\frac{\nabla_{\alpha}(\sqrt{-g}f^{'}g^{\mu\alpha})}{\sqrt{-g}}=-2 f^{'}S^{\mu}
\eeq
which when placed against $(\ref{recal})$ demands that
\beq
S^{\mu}=0
\eeq
and recalling that 
\beq
\frac{\partial_{\mu}f^{'}}{f^{'}}=2\frac{(n-2)}{(n-1)}S_{\mu}
\eeq
it follows that
\beq
\partial_{\mu}f^{'}=0\Rightarrow f^{'}=constant
\eeq
which is true only when $f(R)=R$ and therefore fixing either of $Q_{\mu}$ or $\tilde{Q_{\mu}}$ to zero leads to inconsistency since it forces the $f(R)$ to be linear in $R$. To recap, fixing either $Q_{\mu}=0$ or $\tilde{Q}_{\mu}=0$ in order to break the projective invariance works only for $f(R)=R$ and for general $f(R)$ leads to inconsistencies.\footnote{To be more specific, either of these constraints force the function $f(R)$ to be linear in $R$, which is unreasonable.}  Now, as we have seen fixing $S_{\mu}=0$ breaks the projective invariance and produces a consistent theory. Notice however, that this is not the most general case one can have, especially when one needs to study theories when both the torsion and non-metricity vectors are different from zero. To this end we propose another method that breaks the projective invariance that is more general and instead of setting a vector to zero, establishes a relation between the torsion and non-metricity vectors. We do so in what follows.

\subsubsection{Fixing $(\alpha S_{\mu}-\beta Q_{\mu}-n \gamma \tilde{Q}_{\mu})=0$}
Instead of fixing any of the torsion and non-metricity vectors to zero, here we take a different route and impose a relation between them that can also break the projective invariance. So, what we want to do is take a linear combination of the three vectors that we have and set it to zero, namely
\beq
\alpha S_{\mu}-\beta Q_{\mu}-n \gamma \tilde{Q}_{\mu}=0
\eeq
where $\alpha,\beta,\gamma \neq 0$ are numbers and the minus signs and the factor $n$ are put there just for convenience in the calculation. This constraint is imposed again by means of a Lagrange multiplier 
\beq
S_{A}=\int d^{n}x \sqrt{-g}A^{\mu}(\alpha S_{\mu}-\beta Q_{\mu}-n \gamma \tilde{Q}_{\mu})
\eeq
where $A^{\mu}$ is the Lagrange multiplier that establishes the relation between the three vectors. Our total action is
\begin{gather}
S[g_{\alpha\beta},\Gamma^{\lambda}_{\;\;\;\mu\nu},A_{\rho}]=S_{G}+S_{M}+S_{A}= \\ \nonumber
=\int  d^{n}x\sqrt{-g} \left[ \frac{1}{2\kappa}f(R)+\mathcal{L}_{M}+A^{\mu}(\alpha S_{\mu}-\beta Q_{\mu}-n \gamma \tilde{Q}_{\mu})\right]
\end{gather}
Variation with respect to the Lagrange multiplier gives
\beq
\alpha S_{\mu}-\beta Q_{\mu}-n \gamma \tilde{Q}_{\mu}=0
\eeq
where the parameters $\alpha,\beta,\gamma $ are chosen such as not to preserve the projective invariance. Let us again consider the case where the matter decouples from the connection ($\Delta_{\lambda}^{\;\;\;\mu\nu}=0$) such that $A^{\mu}=0$ and the result after varying with respect to the connection is the same with the one we obtained in the previous subsections, namely
\beq
\frac{1}{2}Q_{\lambda}g^{\mu\nu} -Q_{\lambda}^{\;\;\;\;\mu\nu}-g^{\mu\nu}\frac{\partial_{\lambda}f^{'}}{f^{'}}          +2(S_{\lambda}g^{\mu\nu}+\frac{1}{1-n}S^{\mu}\delta_{\lambda}^{\nu}-  S_{\lambda}^{\;\;\;\mu\nu})=0 \label{kku}
\eeq
\beq
\frac{(n-2)}{2}Q_{\lambda}-n\frac{\partial_{\lambda}f^{'}}{f^{'}}+\frac{2 n(n-2)}{(n-1)}S_{\lambda}=0 \label{eqw1}
\eeq
\beq
-\frac{1}{2}Q^{\mu}+\tilde{Q}^{\mu}+\frac{\partial^{\mu} f^{'}}{f^{'}}-\frac{2(n-2)}{(n-1)}S^{\mu}=0
\eeq
and
\beq
Q_{\mu}-n\tilde{Q}_{\mu}=0
\eeq
Substituting this last equation into the constraint we get
\beq
S_{\mu}=\left( \frac{\beta +\gamma}{\alpha} \right)Q_{\mu}=\lambda Q_{\mu}
\eeq
where we have defined $\lambda=(\beta +\gamma)/\alpha$ and in order to brake the projective invariance it must hold that $\lambda \neq \frac{n-1}{4n}$.\footnote{For this value of the parameter $\lambda$ the combination $S_{\mu}-\lambda Q_{\mu}$ becomes projective invariant.} Now, after some straightforward manipulations of the above equations, one can show that
\beq
S_{\mu}=\lambda Q_{\mu}=\lambda n\tilde{Q}_{\mu}=a\frac{2 n \lambda}{(n-2)} \frac{\partial_{\mu}f^{'}}{f^{'}}
\eeq
where
$a=\frac{1}{1+\frac{4 n}{n-1}}$
From which we see that all three vectors are related to each other and their source is the term $\frac{\partial_{\mu}f^{'}}{f^{'}}$. To gain more intuition on the above, let us vary the total action with respect to the metric tensor to obtain the field equations
\beq
f^{'}(R)R_{(\mu\nu)}-\frac{f(R)}{2}g_{\mu\nu}=\kappa  T_{\mu\nu} \label{emde}
\eeq
where we have also used the fact that $A_{\mu}=0$. Again, taking the trace of the above field equations it follows that
\beq
f^{'}(R)R-\frac{n}{2}f(R)=\kappa T
\eeq
which, as we have already discussed, defines the implicit function $R=R(T)$ and therefore  both $f(R)$ and $f^{'}(R)$  are functions of $T$  ($f(R)=f(R(T))=f(T)$ and $f^{'}(R)=f^{'}(R(T))=f^{'}(T)$). Therefore, a given $T_{\mu\nu}$ will give rise to torsion and non-metricity through its trace and the torsion and non-metricity vectors are related and are proportional to this source which is a function of $T$, that is
\beq
S_{\mu}=\lambda Q_{\mu}=\lambda n\tilde{Q}_{\mu}=a\frac{2 n \lambda}{(n-2)} \frac{\partial_{\mu}f^{'}(T)}{f^{'}(T)}
\eeq
We would now wish to solve explicitly for the torsion and non-metricity tensors and find their exact forms. To do so, we substitute the above relation into
\beq
\frac{1}{2}Q_{\lambda}g^{\mu\nu} -Q_{\lambda}^{\;\;\;\;\mu\nu}-g^{\mu\nu}\frac{\partial_{\lambda}f^{'}}{f^{'}}          +2(S_{\lambda}g^{\mu\nu}+\frac{1}{1-n}S^{\mu}\delta_{\lambda}^{\nu}-  S_{\lambda}^{\;\;\;\mu\nu})=0 \label{kku}
\eeq
to obtain
\beq
(Q_{\lambda}^{\;\;\;\;\mu\nu}+2 S_{\lambda}^{\;\;\;\mu\nu})=b g^{\mu\nu}Q_{\lambda}+\frac{2 \lambda }{1-n}Q^{\mu}\delta_{\lambda}^{\nu} 
\eeq
or
\beq
(Q_{\alpha\mu\nu}+2 S_{\alpha\mu\nu})=b Q_{\alpha} g_{\mu\nu}+\frac{2 \lambda }{1-n}Q_{\mu}g_{\nu\alpha} \label{qstn}
\eeq
where $b=\frac{1}{ n}+\frac{2\lambda}{n-1}$. Note now that this tensor combination along with some index permutations of it appears in the connection decomposition
\begin{equation}
\Gamma^{\lambda}_{\;\;\;\mu\nu}=\tilde{\Gamma}^{\lambda}_{\;\;\;\mu\nu}+\frac{1}{2}g^{\alpha\lambda}\Big( (Q_{\mu\nu\alpha}+2 S_{\mu\nu\alpha})+(Q_{\nu\alpha\mu}+2 S_{\nu\alpha\mu})-(Q_{\alpha\mu\nu}+2 S_{\alpha\mu\nu}) \Big) \label{solk}
\end{equation}
So, carrying out the calculations we finally arrive at
\beq
\Gamma^{\lambda}_{\;\;\;\mu\nu}=\tilde{\Gamma}^{\lambda}_{\;\;\;\mu\nu}+\frac{1}{2}g^{\alpha\lambda}\Big( A(Q_{\mu}g_{\alpha\nu}-Q_{\alpha}g_{\mu\nu})+B Q_{\nu}g_{\mu\alpha}  \Big)
\eeq
where $A=b-\frac{2 n}{n-1}\lambda$,\; $B=b+\frac{2 n}{n-1}\lambda$. Having this one can easily compute the torsion tensor
\beq
S_{\mu\nu}^{\;\;\;\;\lambda}=\Gamma^{\lambda}_{\;\;\;[\mu\nu]}=\frac{2}{n-1}\lambda Q_{[\mu}\delta_{\nu]}^{\lambda}
\eeq
and using $S_{\mu}=\lambda Q_{\mu}$ we also make the consistency check
\beq
S_{\mu\nu}^{\;\;\;\;\lambda}=\frac{2}{n-1} S_{[\mu}\delta_{\nu]}^{\lambda}
\eeq
So, we have the case of a vectorial torsion. As far as non-metricity is concerned, we substitute the last equation into $(\ref{qstn})$ and after some straightforward calculations we finally arrive at
\beq
Q_{\alpha\mu\nu}=\frac{Q_{\alpha}}{n}g_{\mu\nu}
\eeq
which is the case of a Weyl non-metricity. Note that the parameter $\lambda$ has canceled out in the expression for non-metricity. To conclude, what we have done here is to break the projective invariance and produce a viable metric affine $f(R)$ theory. Instead of  setting $S_{\mu}=0$ or $Q_{\mu}=0$ (or even $\tilde{Q}_{\mu}=0$) which singles out a vector out of  the three that are available and therefore constricts the generality, we took a different route  and imposed a constraint on the three vectors ($\alpha S_{\mu}-\beta Q_{\mu}-n \gamma \tilde{Q}_{\mu}=0$) that treats them on equal footing. Our result (when the connection decouples from the matter fields) is a fully consistent theory in which there exist both torsion and non-metricity, powered by a single scalar (T) that is sourced by the energy momentum tensor. More specifically, one has a vectorial torsion and a non-metricity of the Weyl type, with
\beq
S_{\mu\nu}^{\;\;\;\;\lambda}=\frac{2}{n-1} S_{[\mu}\delta_{\nu]}^{\lambda}
\eeq
\beq
Q_{\alpha\mu\nu}=\frac{Q_{\alpha}}{n}g_{\mu\nu}
\eeq
\beq
S_{\mu}=\lambda Q_{\mu}=\lambda n\tilde{Q}_{\mu}=a\frac{2 n \lambda}{(n-2)} \frac{\partial_{\mu}f^{'}(T)}{f^{'}(T)}
\eeq
Some comments are now in order. Firstly, notice that in vacuum ($T_{\mu\nu}=0$) both torsion and non-metricity vanish and therefore they are only introduced by matter fields. Secondly, the above expressions for the affine connection and subsequently for torsion and non-metricity, are algebraic ones since on the assumption that matter decouples from the connection ($\Delta_{\alpha\mu\nu}=0$) we have that $T_{\mu\nu}$ is independent of the connection as seen from $(\ref{emhpt})$. So, breaking the invariance this way we see that the simplest forms of torsion and non-metricity can be sourced by the energy momentum tensor alone, and for further degrees of freedom to be excited, a hypermomentum tensor is also needed.

\subsection{Metric-Affine $f(R)$ with projective invariant matter}
Interestingly, if matter fields that respect the projective invariant are added to $f(R)$ we have exactly the case we presented in Theorem-2. Then, applying the results of our second Theorem we immediately get for the affine connection
\begin{gather}
\Gamma^{\lambda}_{\;\;\;\mu\nu}=\tilde{\Gamma}^{\lambda}_{\;\;\;\mu\nu}+\frac{\kappa}{f'}\frac{g^{\lambda\alpha}}{2}(\Delta_{\alpha\mu\nu}-\Delta_{\nu\alpha\mu}-\Delta_{\mu\nu\alpha})+\frac{\kappa}{f'}\frac{g^{\alpha\lambda}}{(n-2)}g_{\nu[\mu}(\Delta_{\alpha]}-\tilde{\Delta}_{\alpha]}) \nonumber \\
+\frac{1}{(n-2)f'}\Big( \delta^{\lambda}_{\nu}\partial_{\mu}f' -g_{\mu\nu} \partial^{\lambda}f' \Big)\;,\;\;\;\;where\;\;\;\;\;f'=f'(T)
\end{gather}  
With the above connection being dynamical when $T_{\mu\nu}$ depends on the connection, and lacking dynamics when the latter is independent of the connection.

\section{Example 3: A Theory with a dynamical connection}
As an application of our third Theorem let us consider the theory
\beq
S[g,\Gamma]=\int d^{4} x \sqrt{-g}\left( \frac{1}{2 \kappa}R+ \frac{\lambda}{2 \kappa} R_{\mu\nu}R^{\mu\nu} \right)
\eeq
where $\lambda$ is a parameter. Notice that there is no motivation behind the choice of this action, we consider it here as a simple example in order to apply our Theorem-3. It is known in the literature (see \cite{vitagliano2010dynamics} for instance) that Theories of the family $f(R,R_{\mu\nu}R^{\mu\nu})$ admit a dynamical connection in general. Therefore, we expect that in the above Theory the connection is dynamical. This can be easily verified by using our third Theorem. To see this, let us vary the above action with respect to the connection, to get
\beq
P_{\lambda}^{\;\;\;\mu\nu}(g)=-2\lambda P_{\lambda}^{\;\;\;\mu\nu}(R)
\eeq
where $P_{\lambda}^{\;\;\;\mu\nu}(g)$ is the usual Palatini tensor computed with respect to the metric and
\beq
 P_{\lambda}^{\;\;\;\mu\nu}(R)\equiv -\frac{\nabla_{\lambda}(\sqrt{-g}R^{\mu\nu})}{\sqrt{-g}}+\frac{\nabla_{\sigma}(\sqrt{-g}R^{\mu\sigma})}{\sqrt{-g}}\delta_{\lambda}^{\nu}+2(R^{\mu\nu}S_{\lambda}-S_{\alpha}R^{\mu\alpha}\delta^{\nu}_{\lambda}+R^{\mu\sigma}S_{\sigma\lambda}^{\;\;\;\nu})
\eeq
Then using the result $(\ref{theo3})$ of our third Theorem, we have
\begin{equation}
\Gamma^{\lambda}_{\;\;\;\mu\nu}=\tilde{\Gamma}^{\lambda}_{\;\;\;\mu\nu}-g^{\lambda\alpha}\lambda\Big(P_{\alpha\mu\nu}(R)-P_{\nu\alpha\mu}(R)-P_{\mu\nu\alpha}(R)\Big)-2 \lambda\frac{g^{\alpha\lambda}}{(n-2)}g_{\nu[\mu} \Big(P_{\alpha]}(R)-\tilde{P}_{\alpha]}(R)\Big)+\frac{1}{2}\delta_{\mu}^{\lambda}\tilde{Q}_{\nu}  
\end{equation}
and the above is a dynamical equation for the connection. This is easily understood by the appearance of the terms such as $\nabla_{\lambda}R^{\mu\nu}$ which contain higher order terms and derivatives of the connection. So, with this simple example we see an immediate application of our third Theorem. It goes beyond the purposes of this letter to investigate the above theory any further but we mention that a similar theory\footnote{The additional piece they added to the Einstein Hilbert part there was $c_{1}R^{(\mu\nu)}R_{(\mu\nu)}+ c_{2}R^{[\mu\nu]}R_{[\mu\nu]}   $.    } was studied in \cite{vitagliano2010dynamics} . In particular it was shown there that for vanishing torsion, the Theory is equivalent to Einstein's Gravity plus a Proca field \cite{vitagliano2010dynamics}. Similar results (again for vanishing torsion) for an action containing the anti-symmetric part of the Ricci tensor and a quadratic non-metricity term were also found in \cite{allemandi2004accelerated}. However, for projective actions of the form $f(R,R_{(\mu\nu)}R^{(\mu\nu)})$  the connection lacks dynamics \cite{vitagliano2010dynamics}. It would therefore be interesting to classify other actions that give similar results and the conditions upon which the connection lacks/gains. These subjects certainly worth further investigation.

\section{Conclusions}
Metric-Affine Theories of Gravity have a nice feature among other modifications of Gravity. The modifications in this case come naturally by extending the geometry to admit both torsion and non-metricity! As a result the underlying Theory is described by a non-Riemannian geometry. The geometrical structure can then be studied once a metric tensor and a connection are given. It is therefore of great importance to be able to find the affine connection for a given Theory. For Palatini $f(R)$ or  Palatini Ricci squared Theories the procedure on how to solve for the affine connection was known in the literature \cite{allemandi2004accelerated,olmo2009dynamical}. In addition, the way to solve for the Affine connection for Theories containing second order invariants of torsion and non-metricity was given in \cite{vitagliano2014role,vitagliano2011dynamics} . For more general cases however such a method was elusive so far. It was the purpose of this paper to prescribe such a procedure for general classes of Metric-Affine Theories. Let us recap what we have done here.

We presented a systematic way to solve for the affine connection in Metric Affine Theories of Gravity. We stated and proved our results as three consecutive Theorems. Solving for the affine connection in these theories is most important since given an affine connection one can immediately compute the torsion and non-metricity tensors (and curvature too) and therefore have a complete knowledge of the underlying geometry. We showed how one can solve for the affine connection when an action linear in the connection is added to the Einstein-Hilbert action. In this case the connection carries no dynamics, as expected and its expression with respect to the metric and the matter fields is an algebraic one. We then generalized the result  and considered  $f(R)$ theories, with an additional part that is again linear in the connection and its partial derivatives. Finally, we solved for the affine connection  when a general action with no restrictions is added to the Einstein-Hilbert action. In this case however, the equation for the connection is not algebraic, but a differential equation in general. 
 Having proved these Theorems we applied each of them to some simple
 theories in order to illustrate how the procedure, for computing the affine connection, works. In particular, we applied our first Theorem to obtain and confirm the results of \cite{d1982gravity,leigh2009torsion,petkou2010torsional}. In this case the additional term is proportional to torsion and therefore falls in the category of Theorem-1. We then went on to discuss projective invariance breaking in Metric-Affine $f(R)$ theories and presented an alternative way to break the projective invariance. In the case where the matter sector respects projective invariance (then no projective breaking is needed), we showed how one can immediately solve for the connection by using the results of our Theorem-2.  Finally, we presented a simple example (where the Ricci squared is added to $S_{EH}$) with dynamical connection by applying the results of our Theorem-3. We also discussed conditions for having a dynamical/non-dynamical connection and mentioned further possible applications of our results.

\section{Acknowledgments}
I would like to thank Tomi S. Koivisto, Anastasios C. Petkou, Christos G. Tsagas and Lavinia Heisenberg for useful discussions and comments. I would also like to thank Vincenzo Vitagliano for some remarks.

\appendix

\section{Properties of the Palatini tensor}
We prove here some basic properties of the Palatini tensor that we have been using throughout this paper. Recalling its definition
\begin{equation}
P_{\lambda}^{\;\;\;\mu\nu}=-\frac{\nabla_{\lambda}(\sqrt{-g}g^{\mu\nu})}{\sqrt{-g}}+\frac{\nabla_{\sigma}(\sqrt{-g}g^{\mu\sigma})\delta^{\nu}_{\lambda}}{\sqrt{-g}} \\
+2(S_{\lambda}g^{\mu\nu}-S^{\mu}\delta_{\lambda}^{\nu}+g^{\mu\sigma}S_{\sigma\lambda}^{\;\;\;\;\nu})  \nonumber
\end{equation}
and contracting in $\mu,\lambda$, immediately follows that
\begin{equation}
P_{\mu}^{\;\;\;\mu\nu}=-\frac{\nabla_{\mu}(\sqrt{-g}g^{\mu\nu})}{\sqrt{-g}}+\frac{\nabla_{\sigma}(\sqrt{-g}g^{\nu\sigma})}{\sqrt{-g}} \\
+2(S^{\nu}-S^{\nu}+0)=0 \Rightarrow \nonumber
\end{equation}
\begin{equation}
P_{\mu}^{\;\;\;\mu\nu}=0
\end{equation}
thus, the Palatini tensor is traceless in first and second index. Contracting now in $\nu,\lambda$ we have
\begin{equation}
P_{\nu}^{\;\;\;\mu\nu}=(n-1)\frac{\nabla_{\sigma}(\sqrt{-g}g^{\mu\sigma})}{\sqrt{-g}}+2(2-n)S^{\mu}
\end{equation}
and upon using 
\begin{equation}
\nabla_{\sigma}g^{\mu\sigma}=\tilde{Q}^{\mu}
\end{equation}
along with
\begin{equation}
\frac{\nabla_{\sigma}\sqrt{-g}}{\sqrt{-g}}=-\frac{1}{2}Q_{\sigma}
\end{equation}
the latter recasts to
\begin{equation}
P_{\nu}^{\;\;\;\mu\nu}=(n-1)\left[ \tilde{Q}^{\mu}-\frac{1}{2}Q^{\mu}\right] +2(2-n)S^{\mu}
\end{equation}
To obtain a third identity, we multiply (and contract) with $g_{\mu\nu}$ and use the above relations for the Weyl and second non-metricity vector, to arrive at
\begin{equation}
g_{\mu\nu}P_{\lambda}^{\;\;\;\mu\nu}=\frac{(n-3)}{2}Q_{\lambda}+\tilde{Q}_{\lambda}+2(n-2)S_{\lambda}
\end{equation}
Now, defining $P^{\mu}\equiv P_{\nu}^{\;\;\;\mu\nu}$ and $\tilde{P}^{\mu}\equiv g_{\alpha\beta}P^{\mu\alpha\beta}$  adding and subtracting the above two, we get
\beq
P^{\mu}+\tilde{P}^{\mu}=n \tilde{Q}^{\mu}-Q^{\mu}
\eeq
and
\beq
P^{\mu}-\tilde{P}^{\mu}=(n-2) (\tilde{Q}^{\mu}-Q^{\mu}-4 S^{\mu})
\eeq
respectively, and notice that both of the above combinations are projective invariant! Another useful relation comes about by taking the antisymmetric part of the Palatini tensor, which is equal to
\begin{equation}
P_{\lambda}^{\;\;\;[\mu\nu]}=2 A^{[\mu}\delta^{\nu]}_{\lambda}+2 g^{\sigma[\mu}S_{\sigma\lambda}^{\;\;\;\;\;\nu]}
\end{equation}
where
\begin{equation}
A^{\mu}=\frac{1}{2}\tilde{Q}^{\mu}-\frac{1}{4}Q^{\mu}-S^{\mu}
\end{equation}
Using the definitions of non-metricity tensor and vectors we can easily express the Palatini tensor in the form
\begin{equation}
P_{\lambda}^{\;\;\;\mu\nu}=\delta^{\nu}_{\lambda}\left( \tilde{Q}^{\mu}-\frac{1}{2}Q^{\mu}-2 S^{\mu} \right) + g^{\mu\nu}\left( \frac{1}{2}Q_{\lambda}+2 S_{\lambda} \right)-( Q_{\lambda}^{\;\;\;\mu\nu}+2 S_{\lambda}^{\;\;\;\;\mu\nu})
\end{equation}
such that
\begin{equation}
P^{\alpha\mu\nu}= g^{\alpha\nu}\left( \tilde{Q}^{\mu}-\frac{1}{2}Q^{\mu}-2 S^{\mu}\right)+ g^{\mu\nu}\left( \frac{1}{2}Q^{\alpha}+2 S^{\alpha} \right)-( Q^{\alpha\mu\nu}+2 S^{\alpha\mu\nu})
\end{equation}
Note now, that the fully antisymmetric part of the Palatini tensor is determined only by the torsion tensor (the non-metricity part drops out)
\begin{equation}
P^{[\alpha\mu\nu]}=-2  S^{[\alpha\mu\nu]}
\end{equation}
In addition, the completely symmetric part of it is solely determined by non-metricity. Indeed, the above can also be written as
\begin{equation}
P^{\alpha\mu\nu}= g^{\alpha\nu} \tilde{Q}^{\mu}+ 2 g^{\nu[\mu}\left( \frac{1}{2}Q^{\alpha]}+2 S^{\alpha]} \right)-( Q^{\alpha\mu\nu}+2 S^{\alpha\mu\nu})
\end{equation}
and by taking the fully symmetric part it follows that
\begin{equation}
P^{(\alpha\mu\nu)}= g^{(\alpha\nu} \tilde{Q}^{\mu)}-Q^{(\alpha\mu\nu)}
\end{equation}

\section{Derivation of Einstein Field Equations in MAG}

Let us start with the Einstein-Hilbert action in $n$-dimensions
\begin{equation}
S_{EH}[g_{\mu\nu},\Gamma^{\lambda}_{\;\;\;\alpha\beta}]=\int d^{n}x\sqrt{-g}R=\int d^{n}x\sqrt{-g}g^{\mu\nu}R_{(\mu\nu)} \label{ei}
\end{equation}
and no matter fields. Here, no a priori relation between the metric tensor $g_{\mu\nu}$ and the connection $\Gamma^{\lambda}_{\;\;\;\alpha\beta}$ has been assumed and the two are seen as independent fields. Varying ($\ref{ei}$) with respect to $g_{\mu\nu}$ and recalling that $R_{\mu\nu}$ is independent of the metric, we derive
\begin{gather}
\delta_{g}S_{EH}=0 \Rightarrow  \nonumber
0=\int d^{n}x\sqrt{-g}\delta g^{\mu\nu}\Big[ R_{(\mu\nu)}-\frac{g_{\mu\nu}}{2}R\Big] \nonumber
\end{gather}
where we have used the identity
\begin{equation}
\delta_{g}\sqrt{-g}=-\frac{\sqrt{-g}}{2}g_{\mu\nu}\delta g^{\mu\nu}
\end{equation}
Now, since the latter must hold for any arbitrary variation $\delta g^{\mu\nu}$, we have
\begin{equation}
R_{(\mu\nu)}-\frac{g_{\mu\nu}}{2}R=0
\end{equation}
We should point out that at this point that we cannot identify the above as the Einstein equations yet since the torsionlessness and metric compatibility conditions have not been assumed. Now, using 
\begin{equation}
\delta_{\Gamma}R^{\mu}_{\;\;\;\nu\sigma\lambda}=\nabla_{\sigma}\delta \Gamma^{\mu}_{\;\;\;\nu\lambda}-\nabla_{\lambda}\delta \Gamma^{\mu}_{\;\;\;\nu\sigma}-2 S_{\sigma\lambda}^{\;\;\;\;\rho}\delta \Gamma^{\mu}_{\;\;\;\nu\rho}
\end{equation}
and varying  ($\ref{ei}$) with respect to the connection we get
\begin{gather}
\delta_{\Gamma}S_{EH}=0 \Rightarrow \\ \nonumber
0=\int d^{n}x \delta\Gamma^{\lambda}_{\;\;\;\mu\nu}\Big[ -\nabla_{\lambda}(\sqrt{-g}g^{\mu\nu})+\nabla_{\sigma}(\sqrt{-g}g^{\mu\sigma})\delta^{\nu}_{\lambda} \\
+2\sqrt{-g}(S_{\lambda}g^{\mu\nu}-S^{\mu}\delta_{\lambda}^{\nu}+g^{\mu\sigma}S_{\sigma\lambda}^{\;\;\;\;\nu})\Big]
\end{gather}
for this to hold true for any arbitrary variation $\delta\Gamma^{\lambda}_{\;\;\;\mu\nu}$ we must have
\begin{equation}
-\nabla_{\lambda}(\sqrt{-g}g^{\mu\nu})+\nabla_{\sigma}(\sqrt{-g}g^{\mu\sigma})\delta^{\nu}_{\lambda} \\
+2\sqrt{-g}(S_{\lambda}g^{\mu\nu}-S^{\mu}\delta_{\lambda}^{\nu}+g^{\mu\sigma}S_{\sigma\lambda}^{\;\;\;\;\nu})=0 \label{nm}
\end{equation}
which is a relation that relates the metric tensor and the connection. It is common in the literature to denote the left hand side of the above equation (divided by $\sqrt{-g}$) as $P_{\lambda}^{\;\;\;\mu\nu}$ and call it the Palatini tensor. In words
\beq
P_{\lambda}^{\;\;\;\mu\nu}=-\frac{\nabla_{\lambda}(\sqrt{-g}g^{\mu\nu})}{\sqrt{-g}}+\frac{\nabla_{\sigma}(\sqrt{-g}g^{\mu\sigma})\delta^{\nu}_{\lambda}}{\sqrt{-g}} \\
+2(S_{\lambda}g^{\mu\nu}-S^{\mu}\delta_{\lambda}^{\nu}+g^{\mu\sigma}S_{\sigma\lambda}^{\;\;\;\;\nu})
\eeq
Note that in the above case (Einstein-Hilbert action with no matter fields) the Palatini tensor vanishes identically. The Palatini tensor has only $n(n^{2}-1)$ instead of $n^{3}$ due to the fact that is traceless 
\begin{equation}
P_{\mu}^{\;\;\;\mu\nu}=0
\end{equation}
as we have already seen. This implies that a vectorial degree of freedom is left unspecified and as a result the connection can only be determined up to a vector. More specifically, we state that equation ($\ref{nm}$) implies that the connection takes the following form
\begin{equation}
\Gamma^{\lambda}_{\;\;\;\mu\nu}= \tilde{\Gamma}^{\lambda}_{\;\;\;\mu\nu} -\frac{2}{(n-1)}S_{\nu}\delta_{\mu}^{\lambda}=\tilde{\Gamma}^{\lambda}_{\;\;\;\mu\nu}+\frac{1}{2 n}\delta_{\mu}^{\lambda}Q_{\nu}
\end{equation}
where $\tilde{\Gamma}^{\lambda}_{\;\;\;\mu\nu}$ is the Levi-Civita connection. To prove that, we start by contracting ($\ref{nm}$) in $\nu$ and $\lambda$ to get
\begin{equation}
S^{\mu}=\frac{(n-1)}{2(n-2)}\frac{\nabla_{\sigma}(\sqrt{-g}g^{\mu\sigma})}{\sqrt{-g}}
\end{equation}
or
\begin{equation}
\nabla_{\sigma}(\sqrt{-g}g^{\mu\sigma})=2\sqrt{-g}\left(\frac{n-2}{n-1}\right)S^{\mu} \label{expan}
\end{equation}
Substituting that very last equation back to ($\ref{nm}$) we obtain
\begin{equation}
-\nabla_{\lambda}(\sqrt{-g}g^{\mu\nu})
+2\sqrt{-g}\left(S_{\lambda}g^{\mu\nu}+\frac{1}{1-n}S^{\mu}\delta_{\lambda}^{\nu}+g^{\mu\sigma}S_{\sigma\lambda}^{\;\;\;\;\nu}\right)=0 \label{nml}
\end{equation}
Playing a bit more, let us contract ($\ref{nml}$) by $g_{\mu\nu}$. We have
\begin{gather}
-n\frac{\nabla_{\lambda}\sqrt{-g}}{\sqrt{-g}}-g_{\mu\nu}\nabla_{\lambda}g^{\mu\nu}+2\frac{n(n-2)}{(n-1)}S_{\lambda}=0
\end{gather}
Using the identity 
\begin{equation}
\frac{\nabla_{\lambda}\sqrt{-g}}{\sqrt{-g}}=\nabla_{\lambda}\ln{\sqrt{-g}}=\frac{1}{2} g^{\mu\nu}\nabla_{\lambda}g_{\mu\nu}=-\frac{1}{2} g_{\mu\nu}\nabla_{\lambda}g^{\mu\nu}=-\frac{1}{2}Q_{\lambda} \label{as}
\end{equation}
the latter recasts to
\begin{equation}
 \label{sq}
\end{equation}
which relates the torsion and Weyl vectors. One can also relate the second non-metricity vector $\tilde{Q}^{\mu}=Q_{\sigma}^{\;\;\;\sigma\mu}=\nabla_{\sigma}g^{\sigma\mu}$ to $S^{\mu}$ and $Q^{\mu}$. To see this, we expand ($\ref{expan}$) and use ($\ref{as}$) to get
\begin{gather}
g^{\mu\sigma}\frac{\nabla_{\sigma}\sqrt{-g}}{\sqrt{-g}}+\nabla_{\sigma}g^{\sigma\mu}=2\frac{(n-2)}{(n-1)}S^{\mu} 
\end{gather}
or
\begin{gather}
-\frac{1}{2}Q_{\sigma}g^{\mu\sigma}+ \tilde{Q}^{\mu}=2\frac{(n-2)}{(n-1)}S^{\mu}  \nonumber \\
\end{gather}
such that
\begin{equation}
\tilde{Q}^{\mu}=\frac{1}{2}Q^{\mu}+2\frac{(n-2)}{(n-1)}S^{\mu}
\end{equation}
Furthermore, using ($\ref{sq}$) we finally arrive at
\begin{equation}
\tilde{Q}^{\mu}=\frac{1}{n}Q^{\mu}=-\frac{4}{(n-1)}S^{\mu}
\end{equation}
Thus, all three vectors $S^{\mu},Q^{\mu}$ and $\tilde{Q}^{\mu}$ are related to one another. Going back to our proof now, we expand the first term in ($\ref{nml}$) and use equation ($\ref{as}$) along with the definition $Q_{\lambda}^{\;\;\;\mu\nu}\equiv +\nabla_{\lambda}g^{\mu\nu}$, to get
\begin{equation}
\frac{1}{2}g_{\mu\nu}Q_{\lambda}-Q_{\lambda}^{\;\;\;\mu\nu}+2 \left(S_{\lambda}g^{\mu\nu}+\frac{1}{1-n}S^{\mu}\delta_{\lambda}^{\nu}+g^{\mu\sigma}S_{\sigma\lambda}^{\;\;\;\;\nu}\right)=0
\end{equation}
Multiplying with $g^{\alpha\lambda}$ it follows that
\begin{equation}
\frac{1}{2}Q^{\alpha}g^{\mu\nu}-Q^{\alpha\mu\nu}+2\Big( g^{\mu\nu}S^{\alpha}+\frac{1}{1-n}S^{\mu}g^{\nu\alpha}\Big)+2 S^{\mu\alpha\nu}=0   \nonumber
\end{equation}
such that
\begin{equation}
Q^{\alpha\mu\nu}+2 S^{\alpha\mu\nu}=\frac{1}{2}g^{\mu\nu}Q^{\alpha}+2\Big( g^{\mu\nu}S^{\alpha}+\frac{1}{1-n}S^{\mu}g^{\nu\alpha}\Big) \label{qs}
\end{equation}
where the antisymmetry of $S^{\mu\alpha\nu}$ in $\mu,\alpha$ has been employed. Now we use the formula we had proved for the connection decomposition and try to pair the various terms in such a way as to be able to use the above equation. Recalling the decomposition,
\begin{equation}
\Gamma^{\lambda}_{\;\;\;\mu\nu}=\tilde{\Gamma}^{\lambda}_{\;\;\;\mu\nu}+\frac{1}{2}g^{\alpha\lambda}(Q_{\mu\nu\alpha}+Q_{\nu\alpha\mu}-Q_{\alpha\mu\nu}) -g^{\alpha\lambda}(S_{\alpha\mu\nu}+S_{\alpha\nu\mu}-S_{\mu\nu\alpha})
\end{equation}
we use the antisymmetry  $S_{\alpha\nu\mu}=-S_{\nu\alpha\mu}$ in order to re-express the latter as
\begin{equation}
\Gamma^{\lambda}_{\;\;\;\mu\nu}=\tilde{\Gamma}^{\lambda}_{\;\;\;\mu\nu}+\frac{1}{2}g^{\alpha\lambda}\Big[ -(Q_{\alpha\mu\nu}+2 S_{\alpha\mu\nu)}+(Q_{\mu\nu\alpha}+2 S_{\mu\nu\alpha})+(Q_{\nu\alpha\mu}+2 S_{\nu\alpha\mu})\Big]
\end{equation}
Now, multiplying ($\ref{qs}$) by $-1$ and adding the results obtained by successively permuting $\mu\rightarrow \nu$, $\nu\rightarrow \alpha$, $\alpha\rightarrow \mu$ we obtain
\begin{gather}
A_{\alpha\mu\nu}\equiv -(Q_{\alpha\mu\nu}+2 S_{\alpha\mu\nu)}+(Q_{\mu\nu\alpha}+2 S_{\mu\nu\alpha})+(Q_{\nu\alpha\mu}+2 S_{\nu\alpha\mu})= \nonumber \\
=-\frac{1}{2}g_{\mu\nu}Q_{\alpha}-2\Big( g_{\mu\nu}S_{\alpha}+\frac{1}{1-n}S_{\mu}g_{\nu\alpha}\Big) \nonumber \\
+\frac{1}{2}g_{\nu\alpha}Q_{\mu}-2\Big( g_{\nu\alpha}S_{\mu}+\frac{1}{1-n}S_{\nu}g_{\alpha\mu}\Big) \nonumber \\
+\frac{1}{2}g_{\alpha\mu}Q_{\nu}-2\Big( g_{\alpha\mu}S_{\nu}+\frac{1}{1-n}S_{\alpha}g_{\mu\nu}\Big)
\end{gather}
Multiplying with $g^{\alpha\lambda}$ and grouping common terms we obtain 
\begin{gather}
g^{\alpha\lambda}A_{\alpha\mu\nu}=g^{\alpha\lambda}\Big[ -(Q_{\alpha\mu\nu}+2 S_{\alpha\mu\nu)}+(Q_{\mu\nu\alpha}+2 S_{\mu\nu\alpha})+(Q_{\nu\alpha\mu}+2 S_{\nu\alpha\mu})\Big]= \nonumber \\
=-\frac{1}{2}g_{\mu\nu}\Big[ \underbrace{Q^{\lambda}+\frac{4 n}{n-1}S^{\lambda}}_{=0} \Big] + \delta_{(\mu}^{\lambda}Q_{\nu)}+\frac{2 n}{(n-1)}S_{\mu}\delta_{\nu}^{\lambda}+ \frac{2(n-2)}{(n-1)}S_{\nu}\delta_{\mu}^{\lambda}= \nonumber \\
=\frac{1}{2}\delta_{\mu}^{\lambda}\Big[Q_{\nu}+\frac{4(n-2)}{(n-1)}S_{\nu} \Big]+\frac{1}{2}\delta_{\nu}^{\lambda}\Big[\underbrace{ Q_{\mu}+\frac{4 n}{(n-1)}S_{\mu}}_{=0} \Big]= \nonumber \\
=\frac{1}{2}\delta_{\mu}^{\lambda}\Big[\underbrace{Q_{\nu}+\frac{4 n}{(n-1)}S_{\nu}}_{=0}-\frac{8}{(n-1)}S_{\nu} \Big] \Rightarrow \nonumber
\end{gather}
such that
\begin{equation}
g^{\alpha\lambda}A_{\alpha\mu\nu}=-\frac{4}{(n-1)}S_{\nu}\delta_{\mu}^{\lambda}=\frac{1}{n}Q_{\nu}\delta_{\mu}^{\lambda}
\end{equation}
where in all steps we have employed equation ($\ref{sq}$). It is worth noting that the coefficients in front of $g_{\mu\nu}$ and $\delta_{\nu}^{\lambda}$ are exactly equal to zero. Substituting this very last equation into the expression for the connection we complete the proof
\begin{equation}
\Gamma^{\lambda}_{\;\;\;\mu\nu}= \tilde{\Gamma}^{\lambda}_{\;\;\;\mu\nu} -\frac{2}{(n-1)}S_{\nu}\delta_{\mu}^{\lambda}=\tilde{\Gamma}^{\lambda}_{\;\;\;\mu\nu}+\frac{1}{2 n}\delta_{\mu}^{\lambda}Q_{\nu} \label{g}
\end{equation}
Therefore, we conclude that indeed the connection is determined only up to an unspecified vectorial degree of freedom. This additional degree of freedom can be removed by means of a projective transformation of the connection
\begin{equation}
\Gamma^{\lambda}_{\;\;\;\mu\nu}\longrightarrow \Gamma^{\lambda}_{\;\;\;\mu\nu}+\delta_{\mu}^{\lambda}\xi_{\nu}
\end{equation}
if $\xi_{\nu}$ is chosen to be equal to -$Q_{\nu}/2n$. In addition, for connections of the form of ($\ref{g}$) only the Levi-Civita part contributes in both the Einstein-Hilbert action and Einstein's equations. Indeed, substituting ($\ref{g}$) in the definition of the Riemann tensor
\begin{equation}
R^{\mu}_{\;\;\;\nu\alpha\beta}:=2\partial_{[\alpha}\Gamma^{\mu}_{\;\;\;|\nu|\beta]}+2\Gamma^{\mu}_{\;\;\;\rho[\alpha}\Gamma^{\rho}_{\;\;\;|\nu|\beta]}
\end{equation}
It can easily be seen that
\begin{equation}
R^{\mu}_{\;\;\;\nu\alpha\beta}=\tilde{R}^{\mu}_{\;\;\;\nu\alpha\beta}+\frac{1}{n}\delta^{\mu}_{\nu}\partial_{[\alpha}Q_{\beta]}=\tilde{R}^{\mu}_{\;\;\;\nu\alpha\beta}+\frac{1}{n}\delta^{\mu}_{\nu}\hat{R}_{\alpha\beta}
\end{equation}
where $\tilde{R}^{\mu}_{\;\;\;\nu\alpha\beta}$ is the part of the Riemann tensor computed for the Levi-Civita connection, namely the Riemannian part while $\delta^{\mu}_{\nu}\partial_{[\alpha}Q_{\beta]}/n$ represents the non-Riemannian contribution. Subsequently, the Ricci tensor is given by
\begin{equation}
R_{\nu\beta}=\tilde{R}_{\nu\beta}+\frac{1}{n}\partial_{[\nu}Q_{\beta]}=\tilde{R}_{\nu\beta}+\frac{1}{n}\hat{R}_{\nu\beta}
\end{equation}
from which we conclude that its symmetric part (which is the one that contributes to Einstein equations\footnote{This is so because the Einstein Hilbert Lagrangian density is proportional to $R=g^{\mu\nu}R_{\mu\nu}=g^{\mu\nu}R_{(\mu\nu)}$ since the metric tensor is symmetric. As a result, the antisymmetric part of $R_{\mu\nu}$ gives no contribution to the equations of motion.}) is purely Riemannian
\begin{equation}
R_{(\nu\beta)}=\tilde{R}_{(\nu\beta)}=\tilde{R}_{\nu\beta}
\end{equation}
As a result
\begin{equation}
R=g^{\mu\nu}R_{\mu\nu}=g^{\mu\nu}\tilde{R}_{\mu\nu}
\end{equation}
and therefore the additional vectorial degree of freedom does not appear in the Einstein equations. Having solved exactly for the connection we can now compute the torsion and non-metricity tensors in closed form in terms of the unspecified torsion vector (or Weyl vector). Indeed, taking the antisymmetric part of ($\ref{g}$) we obtain for the torsion
\begin{equation}
S_{\mu\nu}^{\;\;\;\;\lambda}=\Gamma^{\lambda}_{\;\;\;[\mu\nu]}= \underbrace{\tilde{\Gamma}^{\lambda}_{\;\;\;[\mu\nu]}}_{=0} -\frac{2}{(n-1)}S_{[\nu}\delta_{\mu]}^{\lambda}
\end{equation}
or
\begin{equation}
S_{\mu\nu}^{\;\;\;\;\lambda}=-\frac{2}{(n-1)}S_{[\nu}\delta_{\mu]}^{\lambda}=\frac{1}{n-1}\Big( S_{\mu}\delta_{\nu}^{\lambda}-S_{\nu}\delta_{\mu}^{\lambda} \Big)
\end{equation}
So long as the non-metricity tensor is concerned, by its definition we have
\begin{gather}
Q_{\alpha\mu\nu}=-\partial_{\alpha}g_{\mu\nu}+\Gamma^{\lambda}_{\;\;\;\mu\alpha}g_{\lambda\nu}+\Gamma^{\lambda}_{\;\;\;\nu\alpha}g_{\lambda\mu} = \nonumber \\
=\underbrace{-\partial_{\alpha}g_{\mu\nu}+\tilde{\Gamma}^{\lambda}_{\;\;\;\mu\alpha}g_{\lambda\nu}+\tilde{\Gamma}^{\lambda}_{\;\;\;\nu\alpha}g_{\lambda\mu}}_{=0} +\frac{1}{2 n}(g_{\lambda\nu}\delta^{\lambda}_{\mu}Q_{\alpha}+g_{\lambda\mu}\delta^{\lambda}_{\nu}Q_{\alpha}) = \nonumber \\
=\frac{1}{2 n}(g_{\mu\nu}Q_{\alpha}+g_{\nu\mu}Q_{\alpha})=\frac{1}{n}Q_{\alpha}g_{\mu\nu}  \nonumber
\end{gather}
where in the second line we used the fact that the non-metricity of the Levi-Civita connection is zero. Therefore,
\begin{equation}
Q_{\alpha\mu\nu}=\frac{1}{n}Q_{\alpha}g_{\mu\nu} 
\end{equation}
Thus we see that both the torsion and non-metricity are non vanishing and dependent on an unspecified vectorial degree of freedom. This is a consequence of the projective invariance of the Einstein-Hilbert action (which results in the tracelessness of the Palatini tensor $P_{\mu}^{\;\;\;\mu\nu}=0$). Finally, the field equations take the form
\begin{equation}
\tilde{R}_{\mu\nu}-\frac{1}{2}\tilde{R}g_{\mu\nu}=0
\end{equation}

\bibliographystyle{unsrt}
\bibliography{refs}

\nocite{koivisto2006note}
\nocite{schouten2013ricci}
\nocite{jimenez2016spacetimes}
\nocite{jarv2018nonmetricity}
\nocite{heisenberg2018systematic}
\nocite{cai2016f}
\nocite{klemm2018einstein}
\nocite{iosifidis2019metric}

\end{document}